\documentclass[%
 reprint,
floatfix,
superscriptaddress,
showpacs,preprintnumbers,
 amsmath,amssymb,
 aps,
pra,
]{revtex4-1}
\usepackage{xcolor}
\usepackage{times}
\usepackage{amssymb}
\usepackage[notrig]{physics}
\usepackage{graphicx}
\usepackage{dcolumn}
\usepackage{bm}
\usepackage{epstopdf}
\usepackage{hyperref}

\begin{document}

\title{Pseudogap phase and fractionalization: Predictions for Josephson junction setup}

\author{Anurag Banerjee}
\affiliation{Institut de Physique Th\'eorique, Universit\'e Paris-Saclay, CEA, CNRS,
F-91191 Gif-sur-Yvette, France.}

\author{Alvaro Ferraz}
\affiliation{International Institute of Physics - UFRN, Natal, Brazil.}

\author{Catherine P\'epin}

\affiliation{Institut de Physique Th\'eorique, Universit\'e Paris-Saclay, CEA, CNRS,
F-91191 Gif-sur-Yvette, France.}
\begin{abstract}
The pseudogap regime of the underdoped cuprates arguably
remains one of the most enigmatic phenomena of correlated quantum
matter. Recent theoretical ideas suggest that a pair density wave
(PDW) or a ``fractionalized PDW'' could be a key ingredient for the understanding
of the pseudogap physics. These ideas are to be contrasted
to the scenario where charge density wave order and superconductivity coexist
at low temperatures. In this paper, we present a few
tests to compare the two scenarios in a Josephson junction setup. For a PDW
scenario, we observe a beat-like structure of AC Josephson current.
The additional frequencies for the AC Josephson current appear at
the half-odd integer multiple of the standard Josephson frequency.
We can extract the modulation wavevector of the PDW state by studying
the average Josephson current. Furthermore, the usual sharp Shapiro
steps break down. In contrast, these signatures are absent for the
simple coexistence of orders. Any detection of such signatures in
a similar experimental setup will strongly support the PDW scenario for the pseudogap phase.
\end{abstract}
\maketitle

\section{Introduction}

One of the most unusual features of the cuprates is the proliferation
of quasi-degenerate orders in the underdoped regime near the mysterious
pseudogap (PG) phase~\cite{Alloul89,Alloul91,Warren89}. These include
experimentally established orders, like superconductivity (SC), charge
density wave (CDW), and antiferromagnetism. Additionally, it may also
harbor putative ``hidden'' orders like the pair density wave (PDW)
state, for which experimental evidence is still lacking. Conceptually,
it is natural to advocate that a very unusual interplay of states
is responsible for the formation of the PG~\cite{loret19}. Many
routes are proposed to drive forward this set of ideas.

Firstly there are the proposals of a vestigial order~\cite{fernandes19,Fradkin15,Sachdev13,Allais14a,Allais14c,Chien09}.
In this framework, the system forms all the potentially degenerate
orders. These orders compete within a standard Ginzburg-Landau description,
resulting in some precursor order that can account for the development
of the pseudogap state. For example, in this scenario, the competition
between the SC and CDW state can lead to a precursor state formed
by a long-ranged PDW order~\cite{PDW20_Nature,shi20_PDW,PDWZnImp_21}.

A more unconventional proposal affirms that such appearance of quasi-degenerate
states can lead to an emergent symmetry~\cite{SachdevPNAS18,Sachdev19,nussinov2002_1,zaanen2003,Lee:1998cr,Dai18,Wang15a}.
Concretely, considering only the SC and CDW orders for simplicity,
the corresponding emergent symmetry is the SU(2) group, rotating between
the two states. However, this emergent symmetry is fragile and it
is easily destroyed by a slight tuning of appropriate parameters.
Nevertheless, the idea of an emergent symmetry is the first illustration
of the presence of some entangled states, in this particular case
the SC and CDW states, which is undoubtedly present in the cuprates.
Indeed some proposals have suggested that around optimal doping, the
cuprate superconductors form a maximally entangled state which is
associated with a strong coupling fixed point that can be accessed
within the holographic framework~\cite{hartnoll18book}.

Somewhere in-between the ideas of an ultimately entangled fixed point
and a vestigial order remains yet another proposal~\cite{Chakraborty19,Grandadam19,pepin20,Grandadam20}.
In this approach, at $T^{*}$ the system is ripe to form all possible
particle-particle (PP) and particle-hole pairs that symmetry allows.
This includes PP pairs with zero and finite momentum, particle-hole
(PH) pairs with finite momentum, some of which are magnetically inert
and others active. Some preformed pairs become unstable and fractionalize
into more robust pairs at lower temperatures. A gauge field emerges,
and the corresponding constraint generated by the fluctuations leads
to the opening of a gap and, thus, to the pseudogap phase itself.
The rest of the introduction highlights the critical differences between
fractionalized PDW and coexisting order scenarios.

\subsection{Theoretical concepts for \textquotedblleft fractionalized\textquotedblright{}
PDW and coexisting orders \label{subsec:Theoretical}}

We focus on the idea of a PDW preformed pair, fractionalizing into
a CDW and SC pairs. The choice has the advantage of simplicity since
these two orders are ubiquitously observed~\cite{Hamidian16,Hoffman02,Wise08,Wen2CDW,Wu:2015bt,CDW_PRL21}
inside the PG region. The CDW pair is given by ${\hat{\chi}_{ij}=g_{\chi}\hat{d}_{ij}\sum_{\sigma}c_{\mathbf{i},\sigma}^{\dagger}c_{j,\sigma}}e^{i\mathbf{Q}_{0}.(\mathbf{r_{i}}+\mathbf{r_{j}})/2}$
where $i,j$ are the nearest neighbor bonds with $\mathbf{Q}_{0}$
being the modulation wave vector, and $g_{\chi}$ the interaction
responsible for forming CDW pairs. $\hat{d}_{ij}$ is the form factor
with a d-wave symmetry. Similarly, the SC pairs is defined as ${\hat{\Delta}_{ij}=g_{\Delta}\hat{d}_{ij}\sum_{\sigma}\sigma c_{i,\sigma}c_{j,-\sigma}}$,
where $c^{\dagger}$ ($c$) are the standard creation (annihilation)
operators for electrons and $g_{\Delta}$ is the interaction forming
the Cooper pairs. The origin of such unstable boson at high temperature
most certainly comes from the strong coupling regime of the electrons~\cite{Baskaran88,Lee92,yang2009nature}
but this is not the main focus of the paper. The PDW is defined as
$\hat{\Delta}_{{\rm PDW}}=g_{{\rm PDW}} c_{i\downarrow}c_{i\uparrow}e^{i\mathbf{Q}_{0}.(\mathbf{r_{i}}+\mathbf{r_{j}})/2}$.
It turns out that it can be written as a combination of elementary
operators, i.e., $\hat{\Delta}_{{\rm PDW}}^{*}=\frac{g_{{\rm PDW}}}{2g_{\Delta}g_{\chi}}\left[\hat{\Delta}_{ij},\hat{\chi}_{ij}^{*}\right]$
and $\hat{\Delta}_{{\rm PDW}}=\frac{g_{{\rm PDW}}}{2g_{\Delta}g_{\chi}}\left[\hat{\chi}_{ij},\hat{\Delta}_{ij}^{*}\right]$
where $\left[a,b\right]$ stands for the commutator of the operators
a and b.

The key idea leading to the fractionalization of the PDW is that at
the PG temperature $T^{*}$ a gauge field (or, equivalently, a local phase) emerges
\begin{align}
\hat{\Delta}_{ij} & \rightarrow\hat{\Delta}_{r}e^{i\theta_{ij}},\nonumber \\
\hat{\chi}_{ij} & \rightarrow\hat{\chi}_{r}e^{i\theta_{ij}},\label{eq:1}
\end{align}
and $\hat{\Delta}_{{\rm PDW}}$ and $\hat{\Delta}_{{\rm PDW}}^{*}$
remain invariant under these transformations. Fluctuations of the gauge field in an effective
field theory generates a constraint (note that the $\Delta$ and $\chi$
have the dimension of energy) 
\begin{align}
\left|\Delta_{r}\right|^{2}+\left|\chi_{r}\right|^{2} & =\left(E^{*}\right)^{2},\label{eq:2}
\end{align}
where $E^{*}$ is an energy scale typical of the PG, which is constant
in temperature, and with respect to spatial variations, but doping
dependent. Note that $\Delta_{r}$ ($\chi_{r}$) are
the local expectation value of the $\hat{\Delta}$ ($\hat{\chi}$)
at position $r$. When the coupling to the conduction electrons
is considered, the constraint of Eq.(\ref{eq:2}) opens a gap, primarily
in the anti-nodal (AN) region of the Fermi surface, leading to the
presence of Fermi arcs in the nodal region~\cite{Chakraborty19,Grandadam19,pepin20,Grandadam20}.
In the temperature regime, $T^{*}>T>T_{co}$, the spatial expectation
values of $\langle\Delta_{r}\rangle=0$ and $\langle\chi_{r}\rangle=0$
as well as the relative phase remains fluctuating.

The typical effective field theory describing the $\hat{\Delta}_{{\rm PDW}}$-
mode, has the form of a quantum rotor model~\cite{Chakraborty19}
\begin{align}
S & =\frac{1}{2}\int d^{2}x\sum_{a,b=1}^{2}\left|\omega_{ab}\right|^{2},\nonumber \\
\mbox{with } & \omega_{ab}=z_{a}\partial_{\mu}z_{b}-z_{b}\partial_{\mu}z_{a},\label{eq:rotor1}
\end{align}
with $z_{1}=\Delta/E^{*}$, $z_{2}=\chi/E^{*}$, $z_{1}^{*}=\Delta^{*}/E^{*}$,
$z_{2}^{*}=\chi^{*}/E^{*}$. Note that $E^{*}$ is a real quantity
associated with the PG energy scale. The gauge fluctuations within the
transformation $z_{a}\rightarrow z_{a}e^{i\theta}$, ($z_{a}^{*}\rightarrow z_{a}^{*}e^{-i\theta}$)
are naturally described by the constraint $\sum_{a}\left|z_{a}\right|^{2}=C$,
(where $C$ is a constant), equivalent to Eq.(\ref{eq:2}). The
model Eq.(\ref{eq:rotor1}) is formally equivalent to the $\mbox{CP}^{1}$
model 
\begin{align}
S= & \int d^{2}x\left|D_{\mu}\psi\right|^{2},\label{eq:6-2}\\
\mbox{with } & D_{\mu}=\partial_{\mu}-i\alpha_{\mu},\nonumber \\
\mbox{and } & \alpha_{\mu}=\frac{1}{2}\sum_{a}z_{a}^{*}\partial_{\mu}z_{a}-z_{a}\partial_{\mu}z_{a}^{*},\nonumber 
\end{align}
with $\psi=\left(z_{1},z_{2}\right)^{T}$~\cite{Perelomov81}.
The model in Eq.(\ref{eq:6-2}) is, in turn, almost the same as a
non-linear $\sigma$-model, but with an additional gauge field $\alpha$
taking care of the intrinsic U(1) gauge symmetry. 

By contrast in the model of coexisting phases the action takes the
form of a standard $\varphi^{4}$-field theory 
\begin{align}
S & =\frac{1}{2}\int d^{2}x\sum_{a=1}^{2}\left(\left|\partial_{\mu}z_{a}\right|^{2}+\mu_{a}\left|z_{a}\right|^{2}\right)\nonumber \\
 & +\sum_{a,b=1}^{2}g_{ab}\left|z_{a}\right|^{2}\left|z_{b}\right|^{2}.\label{eq:m4}
\end{align}
In Eq.(\ref{eq:m4}) the two modes are in coexistence and interact
with each other, but there is no form of chirality and no emerging
gauge field as in Eq.~(\ref{eq:rotor1}).

\subsection{Difference between fractionalized PDW and coexistence}

\label{Intro:Differ} We focus on the underdoped cuprate superconductors
within the two scenarios mentioned above for the PG~\cite{Alloul89,Alloul91,Warren89}.
However, note that the reality is undoubtedly much more complicated
than that due to the proximity to the Mott transition. Nevertheless,
here, we restrict ourselves to discuss these two sets of ideas and
this simplifies the following analysis enormously.

In the coexistence scenario, a precursor of the SC or the CDW orders
forms at $T^{*}$. Most of the theories~\cite{Fradkin15,agterberg2020physics}
that have been advanced so far consider the PDW as a precursor state
at $T^{*}$. Variational calculations on the $tJ$-model show that
the ground state energy of the PDW state is slightly higher than the
ground state energy associated with the uniform superconductivity
regime~\cite{Corboz:2014ba,Choubey17}. Thus, the uniform PDW state
is very difficult to stabilize at low temperatures. Some studies~\cite{agterberg2020physics}
suggest that a locally fluctuating PDW state is responsible for opening
up a pseudogap in the anti-nodal regions without ever being the true
ground state. At a lower temperature $T_{co}$, two-dimensional charge
modulations do form~\cite{Doiron-Leyraud07,Blanco-Canosa13,Sebastian12,Chang16,Wu11}
but they have a short ranged nature. A genuine three-dimensional charge
order only forms under an applied magnetic field~\cite{Gerber:2015gx,Chang16,2016NatComm711747M}.
However, at zero fields, the phase of the CDW is fluctuating rapidly
in space and this produces a very inhomogeneous density-wave pattern,
such as, $\chi\cos\left(\mathbf{Q}_{0}\cdot\mathbf{r}+\phi_{r}\right)$
with $\phi_{r}$ strongly varying from site to site~\cite{Torchinsky:2013dr,Nie13,Nie15,Banerjee18,Campi15}.
The SC order forms at $T_{c}$, leading to the freezing of the SC
phase $\Delta\exp\left(i\theta_{EM}\right)$, due
to the usual Meissner effect.The phase gradient gets minimally coupled
to the external EM vector potential through a term $n_{s}\left(\partial_{\mu}\theta_{EM}-2e\mathbf{A}\right)^{2}$
(where $n_{s}$ is the superfluid density). Then through a gauge transformation,
the vector potential becomes massive, and the SC phase $\theta_{EM}$
gets uniform through the sample. However, in the coexistence scenario,
even below $T_{c}$ the CDW order has no mechanism to have a uniform
phase and should suffer phase fluctuatuations from site to site. Hence,
for the simple coexistence scenario, the entanglement of the PP and
PH pairs is expected to be vanishingly small.

In contrast, in the fractionalization scenario~\cite{Chakraborty19,Grandadam19,pepin20,Grandadam20},
the PDW order fails to form at $T^{*}$. However,
a gauge field (or local phase) emerges, leading to the constraint
of Eq.~(\ref{eq:2}). The difference between the two scenarios sets
in below $T_{c}$, where the two orders coexist. We have now two
orders which are related by an emergent gauge field (or, equivalently, by a local phase
$\theta$) : $\chi\exp\left(i\theta\right)$ and $\Delta\exp\left(i\left(\theta+\theta_{EM}\right)\right)$.
The local phase $\theta$ is minimally coupled to a neutral vector
potential through the term
$n_{\chi}\left(\partial_{\mu}\theta- e^* \mathbf{a}\right)^{2}$ 
(where $n_{\chi}$ is the CDW density and $e^*$ is the fictitious charge). On the other hand,
the SC part has a term $n_{s}\left(\partial_{\mu}\theta+\partial_{\mu}\theta_{EM}-e^*\mathbf{a}-2e\mathbf{A}\right)^{2}$.
When both orders coexist below $T_{c}$ both vector potentials
$\mathbf{a}$ and $\mathbf{a}+\mathbf{A}$ become massive and are
expelled from the sample; in other words, both phases $\theta$
and $\theta_{EM}$ become uniform, and insensitive to impurities~\cite{Chakraborty19}.
Effectively the CDW becomes active to the external EM field and behaves
as a PDW. In the fractionalization scenario, a strong
entanglement thus exists between the PP and PH pairs. Recently
a study reported~\cite{edkins19} an intriguing observation of an
uniform CDW phase inside vortices, and this can be explained by the
fractionalized PDW scenario~\cite{Chakraborty19}. In this study,
we consider another consequence of a fractionalized PDW: namely, its
effects on the current in a Josephson junction setup.

\section{Model and Method}

\label{Sec:Model} 
\begin{figure}[t]
\includegraphics[width=9cm]{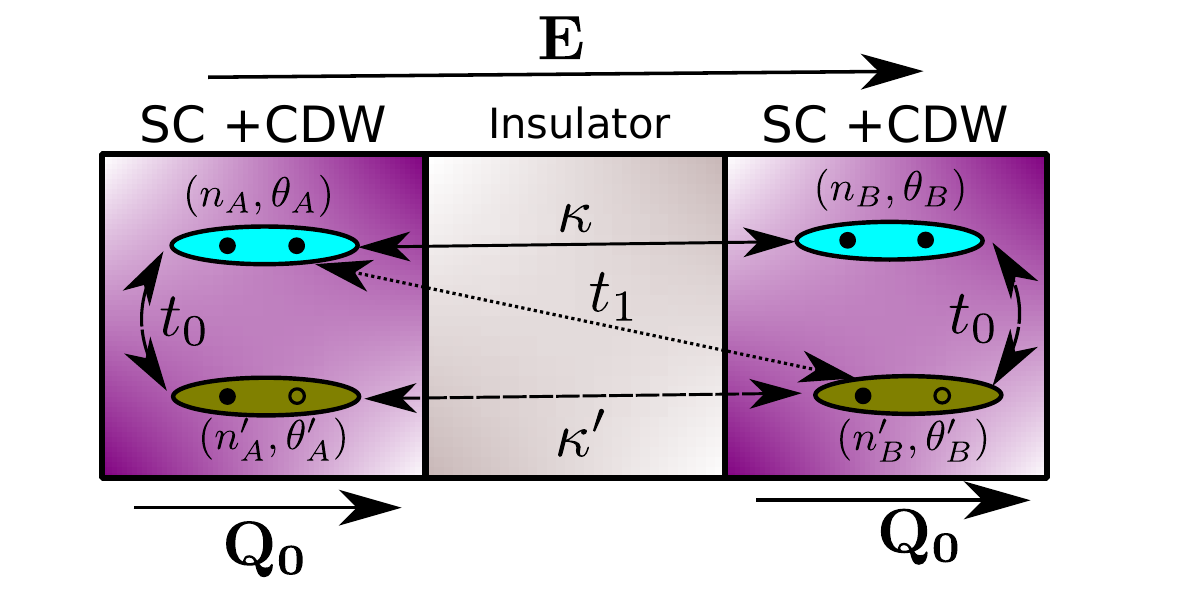}\caption{Josephson junction setup when the applied electric field is parallel
to the charge density wave modulation wavevector $\mathbf{Q}_{0}$.
The different possible hoppings between the pairs are presented schematically.
$\kappa$ denotes a PP-pair hopping from terminal A to B, and vice
versa. $\kappa^{\prime}$ represents the same for the hopping of PH-pairs
across the junction. The inter-junction conversion of PP-pair to a
PH-pair is denoted by $t_{1}$ and the same within the junction by
$t_{0}$.}
\label{Fig:fig1} 
\end{figure}

We propose the experimental setup pictured in Fig.~(\ref{Fig:fig1}).
A Josephson junction (JJ) is considered, with underdoped cuprates
compounds in two terminals A and B separated by an insulating material.
Below $T^{*}$, the terminals in A and B access the PG regime and
we expect these terminals to display both SC and CDW orders at low
temperatures. The JJ is oriented such that the modulation wave vector
is parallel to the junction. Moreover, in the present situation, an
electric field $\mathbf{E}$ is applied in the same direction as CDW
wavevector $\mathbf{Q}_{0}$.

We consider the following wavefunctions for the A and B subsystems
\begin{align}
\ket{\psi_{A}} & =\sqrt{n_{A}}e^{i\theta_{A}}+\sqrt{n_{A}^{\prime}}e^{i\theta_{A}^{\prime}},\nonumber \\
\ket{\psi_{B}} & =\sqrt{n_{B}}e^{i\theta_{B}}+\sqrt{n_{B}^{\prime}}e^{i\theta_{B}^{\prime}},\label{eq:5}
\end{align}

where $\left(n_{A,B},\theta_{A,B}\right)$ are the superfluid density
and phases of SC states on terminals A and B respectively, and $\left(n_{A,B}^{\prime},\theta_{A,B}^{\prime}\right)$
are the corresponding CDW density and phases. At a steady state, we
take $\theta_{A}^{\prime}=\mathbf{Q}_{0}\cdot\mathbf{r}$ and $\theta_{B}^{\prime}=\mathbf{Q}_{0}\cdot\left(\mathbf{r}+\mathbf{\delta}\right)$.
The most generic set of Schr\"{o}dinger's equations~\cite{JJbook,GrunerRMP}
that can be written with these two orders, with $\overline{\psi}=\left(\begin{array}{cccc}
\sqrt{n_{A}}e^{i\theta_{A}}, & \sqrt{n_{B}}e^{i\theta_{B}}, & \sqrt{n_{A}^{\prime}}e^{i\theta_{A}^{\prime}}, & \sqrt{n_{B}^{\prime}}e^{i\theta_{B}^{\prime}}\end{array}\right)$ 
\begin{align}
i\hbar\frac{\partial}{\partial t}\psi & =\left(\begin{array}{cccc}
\mu_{\Delta} & \kappa & t_{0} & t_{1}\\
\kappa & -\mu_{\Delta} & t_{1} & t_{0}\\
t_{0} & t_{1} & V_{A} & \kappa^{\prime}\\
t_{1} & t_{0} & \kappa^{\prime} & V_{B}
\end{array}\right)\psi,\label{eq:6}
\end{align}
where $\kappa$ and $\kappa^{\prime}$ are the hopping integrals for
particle-particle (PP) and particle-hole (PH) pairs across the junctions.
The parameter $\mu_{\Delta}=eU$, where $U$ is the electrical potential
applied to the PP pairs. In terms of the applied electric field $\mathbf{E}$,
this becomes $\mu_{\Delta}=-e\mathbf{E}r_{A,B}$, where $r_{A,B}$
is the distance with respect to the center. However, in this study,
$\mu_{\Delta}$ is assumed to be a constant as in the standard JJ
setup~\cite{JJbook}. The parameter $t_{1}$ represents the tunneling
between the PP pairs in the junction A (resp. B) to the modulated
PH pairs in junction B (resp. A), whereas $t_{0}$ is the same type
of tunneling within a junction. The electric field acts on the CDW
sector~\cite{leeRice79} as $V_{A,B}=eE\bar{\rho}\theta_{A,B}^{\prime}$,
where $\bar{\rho}=n^{\prime}/Q_{0}$ is an effective CDW density divided
by the ordering wave vector $\mathbf{Q}_{0}$. We have assumed that
the superfluid density and the CDW densities are constant throughout
the JJ. Although this is not necessary; such assumption simplifies
the following analysis enormously. Equations (\ref{eq:6}) can be decoupled
into (for details see Appendix~(\ref{AppEqns})) 
\begin{align}
\frac{\partial\bar{n}}{\partial t}=\frac{4}{\hbar} & \left[-\kappa n\sin{\theta}+\sqrt{nn^{\prime}}\cos{\frac{\phi}{2}}\right.\nonumber \\
 & \left.\times\left(t_{0}\sin{\frac{\theta^{\prime}-\theta}{2}}-t_{1}\sin{\frac{\theta^{\prime}+\theta}{2}}\right)\right],\label{eq:m5-1}\\
\frac{\partial\theta}{\partial t}=\frac{2}{\hbar} & \left[\mu_{\Delta}+\sqrt{\frac{n^{\prime}}{n}}\sin{\frac{\phi}{2}}\right.\nonumber \\
 & \left.\times\left(t_{0}\sin{\frac{\theta^{\prime}-\theta}{2}}-t_{1}\sin{\frac{\theta+\theta^{\prime}}{2}}\right)\right],\label{eq:m6-1}\\
\frac{\partial\bar{n}^{\prime}}{\partial t}=\frac{4}{\hbar} & \left[-\kappa^{\prime}n^{\prime}\sin{\theta^{\prime}}+\sqrt{nn^{\prime}}\cos{\frac{\phi}{2}}\right.\nonumber \\
 & \left.\times\left(t_{0}\sin{\frac{\theta-\theta^{\prime}}{2}}-t_{1}\sin{\frac{\theta^{\prime}+\theta}{2}}\right)\right],\label{eq:m7}\\
\frac{\partial\theta^{\prime}}{\partial t}=\frac{1}{\hbar} & \left[(V_{A}-V_{B})-2\sqrt{\frac{n^{\prime}}{n}}\sin{\frac{\phi}{2}}\right.\nonumber \\
 & \left.\times\left(t_{0}\sin{\frac{\theta-\theta^{\prime}}{2}}-t_{1}\sin{\frac{\theta^{\prime}+\theta}{2}}\right)\right],\label{eq:m7-1}\\
\frac{\partial\phi}{\partial t}= & -\frac{1}{\hbar}\left[(V_{A}+V_{B})+2\kappa\cos{\theta}+2\kappa^{\prime}\cos{\theta^{\prime}}\right.\nonumber \\
 & \left.-\frac{2\delta n}{\sqrt{nn^{\prime}}}\left(t_{0}\cos{\frac{\theta-\theta^{\prime}}{2}}+t_{1}\cos{\frac{\theta+\theta^{\prime}}{2}}\right)\right],\label{eq:m8}
\end{align}
with $\theta=\theta_{B}-\theta_{A}$, $\theta^{\prime}=\theta_{B}^{\prime}-\theta_{A}^{\prime}$
,$\phi=\theta_{B}^{\prime}+\theta_{A}^{\prime}-\theta_{A}-\theta_{B}$,
$\bar{n}=n_{B}-n_{A}$, $\bar{n}^{\prime}=n_{A}^{\prime}-n_{B}^{\prime}$,
$\delta n=n^{\prime}-n$. Furthermore, we assumed that the density
of PP-pairs and PH-pairs in both the terminal are similar, i.e. $n\simeq n_{A}\simeq n_{B}$,
$n^{\prime}\simeq n_{A}^{\prime}\simeq n_{B}^{\prime}$. The variation
of $\phi$ can be simplified further in the limit where $n\simeq n^{\prime}$,
$\delta n\ll n$. In this limit, the Eq.~(\ref{eq:m8}) simplifies
to, 
\begin{align}
\frac{\partial\phi}{\partial t}\approx & -\frac{1}{\hbar}\left[(V_{A}+V_{B})+2\kappa\cos{\theta}+2\kappa^{\prime}\cos{\theta^{\prime}}\right].\label{eq:m8-1}
\end{align}
Next we elucidate on the other parameters. The difference of the electric
field on the CDW sector can be simplified to ${V_{A}-V_{B}=-(eEn^{\prime}/Q_{0})\theta^{\prime}\equiv-r_{0}\theta^{\prime}}$,
where we have introduced a new parameter $r_{0}=eEn^{\prime}/Q_{0}$.
Similarly, the average potential acting on the CDW sector due to the
applied electric field can be treated as constant, i.e. $V_{A}+V_{B}=\eta$.

\begin{figure}[t]
\includegraphics[width=9cm]{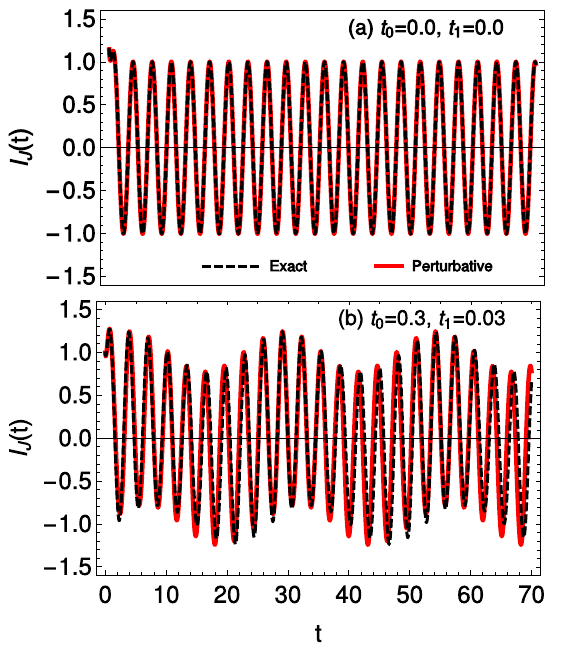} \caption{The variation of Josephson current with time for the setup presented
in Fig.~(\ref{Fig:fig1}), i.e., the applied electric field is parallel
to the charge modulation wavevector. Here we compare the exact numerical
results with the analytical form of the current calculated perturbatively.
\textbf{(a)} Shows the situation when the two orders coexist, for
a vanishing $t_{0}$ and $t_{1}$. In this situation, the perturbative
calculations are exact. The initial transient current regime for CDW
order vanishes at a long-time limit, and we are left with the simple
ac-Josephson effect. \textbf{(b)} Presents the Josephson current for
the situation when the entanglement between the orders forms a fractionalized
PDW or PDW with a finite $t_{0}=0.3,t_{1}=0.03$. The perturbative
calculations and the numerical analysis shows a good match. Interestingly,
the presence of entanglement between CDW and SC orders induces a beat-like
modulation of the Josephson current, which can be contrasted with
the simple coexistence scenario presented above. }
\label{Fig:fig2} 
\end{figure}

\section{Results}

In this section we provide the results by solving the Eqs.~(\ref{eq:m5-1}-\ref{eq:m8-1})
and finding the Josephson current. First we discuss the terms $t_{0}$
and $t_{1}$ that generates the main difference between the two scenarios.
For the simple coexistence of orders, the phase of
the CDW and SC orders are not linked as discussed in the Sec.~(\ref{Intro:Differ}).
Therefore, CDW pairs have no mechanism for having a uniform phase
over the whole sample. Consequently, it suffers phase shifts and will
fluctuate widely from site to site. Such an incoherent
CDW pattern is expected to have minimal overlap with the SC pairs.
 Therefore, we can safely ignore the hopping from a PP-pair to a
PH-pair, as the quantum entanglement between the orders is weak. We
can set the parameters ${t_{0}=\braket{\psi_{SC}^{A\left(B\right)}\vert\psi_{CDW}^{A\left(B\right)}}=0}$
and ${t_{1}=\braket{\psi_{SC}^{A\left(B\right)}\vert\psi_{CDW}^{B\left(A\right)}}=0}$
in Eq.(\ref{eq:6}) for the coexistence of orders.

In the fractionalized PDW case the CDW phase $\theta^{\prime}$ is
such that phase of the two electrodes is given by 
\begin{align}
\theta_{A}^{\prime} & =\mathbf{Q}_{0}\cdot\mathbf{x}+\delta\theta_{A}^{\prime},\nonumber \\
\theta_{B}^{\prime} & =\mathbf{Q}_{0}\cdot\mathbf{\left(x+\delta\right)}+\delta\theta_{B}^{\prime},\label{eq:m11}
\end{align}
where $\delta$ is a dephasing from electrode A to B. $\delta\theta_{A(B)}^{\prime}$
are the fluctuations of the phase from the initial steady state. In
the setup of Fig.~(\ref{Fig:fig1}) when the electric field and wavevector
$Q_{0}$ are parallel, $\delta$ is controlled by distance between
the two electrodes. In Fig.~(\ref{Fig:fig5}a) we study an independent
configuration, where the electric field and $Q_{0}$ are perpendicular.
In such a case the dephasing is controlled by the initial the phase-shift
of the CDW wavevector in the direction perpendicular to the electrodes.
The main difference between the coexisting case is that now $\theta$
and $\theta^{\prime}$ are not independent. Since the phase of $\Delta_{{\rm PDW}}$
is coupled to the EM field but has $\mathbf{Q}_{0}$ modulations as
well, we have 
\begin{align}
\theta & =\delta\theta^{\prime}+\theta_{EM},\label{eq:m12}
\end{align}
where $\theta_{EM}$ is the electromagnetic phase and $\delta\theta^{\prime}=\delta\theta_{B}-\delta\theta_{A}$.
Assuming that we are deep inside the SC phase, with un-fractionalized
Cooper pairs around, the Meissner effect leads to a uniform EM field~\cite{Chakraborty19}.
Thus effectively the CDW becomes active to the external EM field,
due to the Meissner effect as the phase of the CDW becomes insensitive
to the impurities and acts like a PDW order. Therefore a uniform CDW
pattern induces inside the SC phase, an astonishing result that has
been reported in Scanning Tunneling Microscopy (STM) experiment~\cite{edkins19}.
Since all the phases are fixed, we now get a finite tunneling $t_{1}$
and $t_{0}$ back and forth from the modulated CDW and SC phases,
and Eqs.~(\ref{eq:m5-1}-\ref{eq:m8-1}) needs to be solved with
$t,t_{0}\neq0$.

For both the scenario, the usual Josephson current is modified due
to the presence of the CDW~\cite{JJbook,GrunerRMP}, with 
\begin{align}
I_{J} & =-\frac{1}{2}\frac{\partial\bar{n}}{\partial t}-\frac{1}{\pi}\frac{\partial\theta^{\prime}}{\partial t}.\label{eq:m9}
\end{align}
Note that the two fluids couple in the opposite way to the field --
the tunneling of the charge two bosons contributes to the conventional
SC Josephson current. In contrast, the variation of the phase creates
a charge imbalance in the case of the CDW and generates an additional
current. We solve the Eqs.~(\ref{eq:m5-1}-\ref{eq:m8-1}) and evaluate
the Josephson current by using Eq.~(\ref{eq:m9}). Since the difference
between the two scenarions sets in at $T_{c}$, we focus on on the
zero temperature limit of the Josephson current. Next we study, the
alternating current (AC) Josephson effect by applying a constant potential
difference between the terminals.

\subsection{AC Josephson effect}

\subsubsection{Coexistence case}

For the simple coexistence of orders since the parameters $t_{0}$
and $t_{1}$ vanishes, the Eqns.~(\ref{eq:m5-1}-\ref{eq:m8-1})
simplifies enormously. Solving for Eq.(\ref{eq:m9}) we get 
\begin{align}
I_{J} & =\frac{2\kappa n}{\hbar}\sin\left(\frac{2\mu_{\Delta}}{\hbar}t+C_{1}\right)+\frac{C_{2}r_{0}}{\pi\hbar}e^{-r_{0}t}.\label{eq:mcur_coex}
\end{align}
Here $C_{1}$ and $C_{2}$ are constants of integration and depends
on the initial experimental setup. The first term in Eq.(\ref{eq:mcur_coex})
is the standard AC Josephson current with the primary Josephson frequency,
$\omega_{0}=(2eU)/\hbar$. Similarly, one can define standard Josephson
time-period by $T=2\pi/\omega_{0}$. The second term arises due to
CDW in the situation of coexisting order. Since the SC and CDW orders
are disconnected, we only observe a transient response from the CDW
phase variation. We have plotted the current $I_{J}$ as a function
of time $t$ in Fig.~(\ref{Fig:fig2}a) for the coexistence of CDW
and SC. The parameters used to obtain Fig.~(\ref{Fig:fig2}a) are
given by $\mu_{\Delta}=1.0$, $n=1.0$, $n^{\prime}=1.0$, $r_{0}=3.0$,
$\eta=0.3$, $\kappa=0.5$, $\kappa^{\prime}=0.6$ with $t_{1}=t_{0}=0$.
We measure all the energies in the units of $2\kappa$ which we have
set to unity. Furthermore, we have also set the constants $e=1$ and
$\hbar=1$. We have also used the initial conditions, $\theta(t=0)=0$,
$\phi(0)=0$, and $\theta^{\prime}(0)=0.6$. The transient CDW regime
near $t=0$ paves the way to a standard form of the Josephson current
for an standard SC current for $t\gg r_{0}$. In this situation, $I_{J}(t)$
from the numerical calculations presented in black dotted trace and
the analytical form of Eq.~(\ref{eq:mcur_coex}) depicted in red
thick trace in Fig.~(\ref{Fig:fig2}a) matches exactly.

\subsubsection{Fractionalized PDW case}

In the fractionalized PDW scenarion the terms $t_{0}$ and $t_{1}$
are non-vanishing. We have obtained an approximate solution of the
Josephson current in Appendix~(\ref{AppJosephCurr}). The Josephson
current in the first order perturbation in $t_{0}$ and $t_{1}$ is
given by 
\begin{align}
I_{J}= & \frac{2\kappa n}{\hbar}\sin\left(\frac{2\mu_{\Delta}}{\hbar}t+C_{1}\right)+\frac{C_{2}r_{0}}{\pi\hbar}e^{-r_{0}t}\nonumber \\
 & -\frac{2\sqrt{nn^{\prime}}}{\hbar}\cos{\frac{\phi_{0}}{2}}\left(t_{0}\sin{\frac{\theta_{0}^{\prime}-\theta_{0}}{2}}-t_{1}\sin{\frac{\theta_{0}^{\prime}+\theta_{0}}{2}}\right)\nonumber \\
 & +\frac{2}{\pi\hbar}\sqrt{\frac{n^{\prime}}{n}}\sin{\frac{\phi_{0}}{2}}\left(t_{0}\sin{\frac{\theta_{0}-\theta_{0}^{\prime}}{2}}-t_{1}\sin{\frac{\theta_{0}+\theta_{0}^{\prime}}{2}}\right).\label{eq:mCurPDW}
\end{align}
The time evolution of zeroth order($t_{0}=0$, $t_{1}=0$) solutions
of $\theta_{0}$, $\theta_{0}^{\prime}$, and $\phi_{0}$ is given
by 
\begin{align}
\theta_{0} & =\frac{2\mu_{\Delta}}{\hbar}t+C_{1},\label{eq:mtheta0}\\
\theta_{0}^{\prime} & =\frac{C_{2}}{\hbar}e^{-r_{0}t},\label{eq:mtheta0p}\\
\phi_{0} & =-\frac{\eta}{\hbar}t+C_{3}+\frac{2\kappa^{\prime}}{r_{0}\hbar}\text{Ci}\left[C_{2}e^{-r_{0}t}\right]-\frac{\kappa}{\mu_{\Delta}}\sin\left(\frac{2\mu_{\Delta}}{\hbar}t+C_{1}\right),\label{eq:mphi0}
\end{align}
where again $C_{3}$ is a constant of integration and $\text{Ci}[x]$
is the cosine integral function.

We have displayed the current of Eq.~(\ref{eq:mCurPDW}) in Fig.~(\ref{Fig:fig2}b)
for the same set of parameters as in Fig.~(\ref{Fig:fig2}a) albeit
with a finite $t_{0}=0.3,t_{1}=0.03$. The perturbative analytical
calculations matches well with the exact numerical form of the current.
The Josephson current for the fractionalized PDW displays a beat-like
structure which is strikingly distinguishable from the coexistence
case. This provides us with the first prediction -- If the PG phase
of the underdoped cuprates supports a PDW or fractionalized PDW state,
the AC Josephson current should develop a beat-like form as shown
in Fig.~(\ref{Fig:fig2}b). Whereas if the orders simply coexist
in the PG phase the AC Josephson current in long-times will follow
the conventional form.

\subsubsection{Frequency of the AC Josephson current}

\begin{figure*}[t]
\includegraphics[width=16cm]{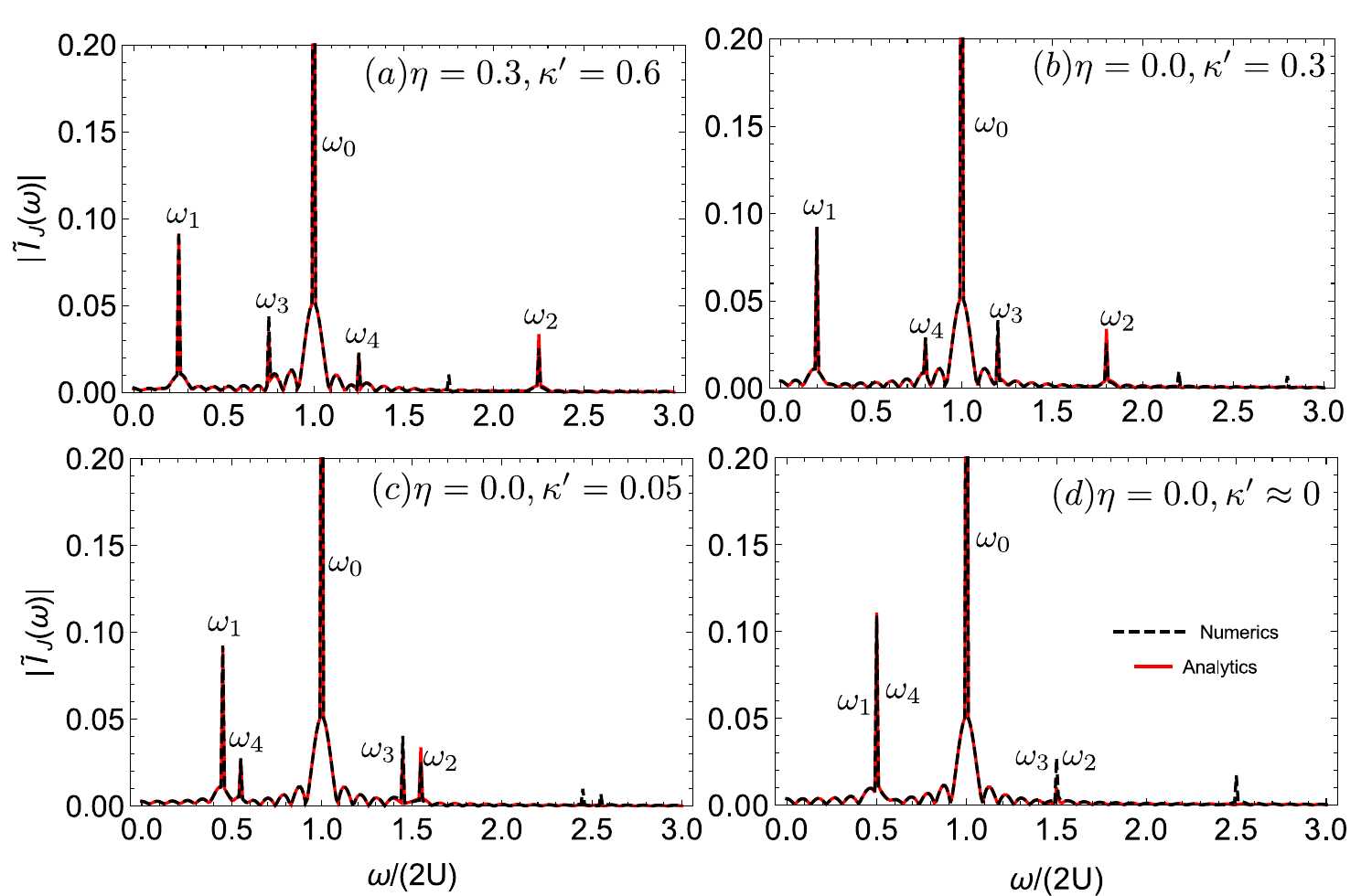} \caption{Exhibits the primary $\omega_{0}$ and additional frequencies $\omega_{i}$
of AC Josephson current for $t_{0}=0.3,t_{1}=0.0$. Here $\kappa^{\prime}$
and $\eta$ are material-dependent parameters. As the potential difference
between the two terminals dominates the material-dependent parameters,
the number of peaks reduces due to the merging of peaks as shown in
\textbf{(c-d)}. The additional peaks for $U\gg\eta+2\kappa^{\prime}$
are at the half-odd integer multiples of the primary peak $\omega_{0}$.}
\label{Fig:fig3} 
\end{figure*}

The Josephson current for a fractionalized PDW state shows a beat-like
structure that suggests multiple frequencies contribute to the AC
Josephson current. To get the frequency, we need to perform a Fourier
transform of the AC Josephson current to the frequency domain.

We simplify the Eq.~(\ref{eq:mCurPDW}) by assuming that the inter
junction PP to PH hopping amplitude is small compared to the intra-junction
hoppings, i.e., $t_{1}\ll t_{0}$. In an experimental scenario, this
requires using a barrier to decay the PH hoppings across the junction.
However, for a finite but small $t_{1}$, will not create any qualitative
difference to the discussion below. Also, in a long time limit, the
transient current regime from the CDW vanishes, i.e., $\theta_{0}^{\prime}\rightarrow0$.
Following the manipulations detailed in Appendix.~(\ref{App:Freq}),
we obtain 
\begin{align}
I_{J}\approx & \frac{t_{0}\sqrt{nn^{\prime}}}{\hbar}\left[\mathcal{J}_{0}(b)\left\lbrace \sin a_{+}t+\sin a_{-}t\right\rbrace +\right.\nonumber \\
 & \left.\mathcal{J}_{1}(b)\left\lbrace \sin\left((a_{+}+\xi)t\right)-\sin\left((a_{-}+\xi)t\right)\right.\right.\nonumber \\
 & \left.\left.+\sin a_{-}t-\sin a_{+}t\right\rbrace \right]+\frac{2\kappa n}{\hbar}\sin{\xi t},\label{eq:mCurrBassel}
\end{align}
where $\mathcal{J}_{\nu}(x)$ is the Bessel function of first kind
of the $\nu$-th order. Also we have redefined 
\begin{align}
a_{\pm} & =\frac{1}{2}\left(\frac{2\mu_{\Delta}}{\hbar}\pm\frac{\eta}{\hbar}\pm\frac{2\kappa^{\prime}}{\hbar}\right),\\
b & =\frac{\kappa}{2\mu_{\Delta}},\\
\xi & =\frac{2\mu_{\Delta}}{\hbar}.
\end{align}
Since, all the terms are directly proportional to $t$, one can easily
read off the frequencies for the AC current, by performing a Fourier
transform. This is given by 
\begin{align}
\tilde{I}_{J}(\omega)= & \frac{\kappa n}{\hbar}\delta({\omega-\xi})+\frac{t_{0}\sqrt{nn^{\prime}}}{2\hbar}\left[(\mathcal{J}_{0}(b)+\mathcal{J}_{1}(b))\delta(\omega-a_{-})\right.\nonumber \\
 & \left.+\mathcal{J}_{1}(b)\delta(\omega-a_{+}-\xi)-\mathcal{J}_{1}(b)\delta(\omega-a_{-}-\xi)\right.\nonumber \\
 & \left.+(\mathcal{J}_{0}(b)-\mathcal{J}_{1}(b))\delta(\omega-a_{-})\right].\label{eq:mfreq}
\end{align}
The primary frequency $\xi$ is the usual AC-Josephson frequency.
The other frequencies are the additional originating due to the entanglement
of the two orders. The ratio of the primary to the few additional
frequency is given by,

\begin{align}
\frac{\omega_{1}}{\omega_{0}}=\frac{1}{2}-\left(\frac{\eta+2\kappa^{\prime}}{4eU}\right),\label{eq:freq1}\\
\frac{\omega_{4}}{\omega_{0}}=\frac{1}{2}+\left(\frac{\eta+2\kappa^{\prime}}{4eU}\right),\label{eq:freq2}\\
\frac{\omega_{2}}{\omega_{0}}=\frac{3}{2}+\left(\frac{\eta+2\kappa^{\prime}}{4eU}\right),\label{eq:freq3}\\
\frac{\omega_{3}}{\omega_{0}}=\frac{3}{2}-\left(\frac{\eta+2\kappa^{\prime}}{4eU}\right),\label{eq:freq4}
\end{align}
where $\omega_{i-1}$ are the $i$-th delta-function peak of Eq.~(\ref{eq:mfreq})
and used the fact that $\mu_{\Delta}=eU$, where $U$ is the DC potential
applied across the terminals. For large potential difference between
the two junctions, i.e., $U\gg\eta+2\kappa^{\prime}$, the second
term of all the ratios vanishes. Moreover, the ratio between the primary
and additional frequencies will occur at half-odd integers, i.e.,
$1/2,3/2,5/2,...$. However, the peak intensity will diminish for
the higher-order peaks as the higher-order Bessel functions determine
their strength.

We have presented in Fig.~(\ref{Fig:fig3}) the Fourier transform
of the Josephson current given in Eq.~(\ref{eq:mCurPDW}) solved
numerically in compared with the perturbative solution of Eq.~(\ref{eq:mfreq}).
The dominant peaks of the AC Josephson junction are captured within
our approximate analysis. The primary peak $\omega_{0}$ arising from
the normal AC Josephson effect remains unchanged for all the parameters.
In Fig.~(\ref{Fig:fig3}a) and Fig.~(\ref{Fig:fig3}b) when $U\sim\eta+2\kappa^{\prime}$,
the four additional peaks are separated and well-resolved. The separation
of the peaks $\omega_{1}$ and $\omega_{4}$ from $0.5\omega_{0}$
(similarly $\omega_{2}$ and $\omega_{3}$ from $1.5\omega_{0}$)
should reduce monotonically as the potential difference between the
two electrodes increases. This is illustrated in Fig.~(\ref{Fig:fig3}b)
to Fig.~(\ref{Fig:fig3}d). When $U\gg\eta+2\kappa^{\prime}$, the
two peaks merge as shown in Fig.~(\ref{Fig:fig3}c) and Fig.~(\ref{Fig:fig3}d)
which leads to an apparent reduction in the number of peaks. Therefore
when the potential difference between the terminals is large compared
to the material-dependent parameters, the additional frequencies occur
at half-odd integers of $\omega_{0}$. Studying such frequency dependence
of the AC Josephson current peaks will give strong evidence for the
fractionalized PDW scenario.

\subsubsection{Envelope of the AC Josephson current}

\begin{figure}[t]
\includegraphics[width=9cm]{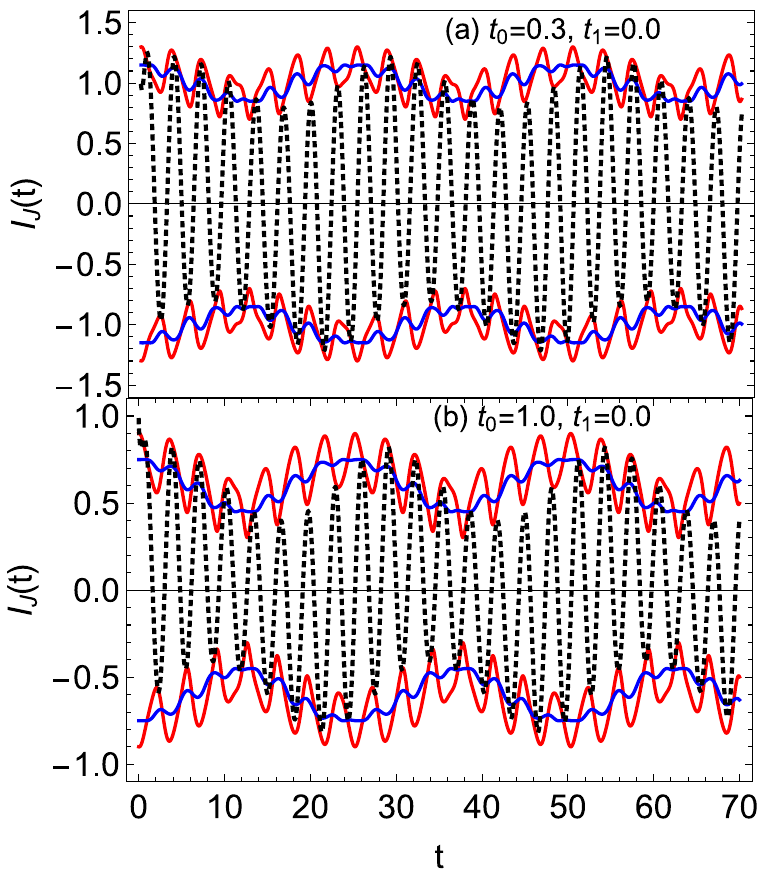} \caption{Presents the evolution of the beat-like form of the Josephson current
for two different parameters \textbf{(a)} $t_{0}=0.3$, \textbf{(b)}
$t_{0}=1$. The total envelope for the oscillation is governed by
the red traces, which still contains some modulations. The blue trace
shows the slower oscillation of the envelope which is controlled by
the first term of Eq.~(\ref{eq:menv}) and hence by $\chi_{1}=(\phi_{0}-\theta_{0})/2$.
This additional oscillation of the envelope is due to the entanglement
between the SC and CDW order.}
\label{Fig:fig4} 
\end{figure}

We have also obtained the envelope for the oscillation observed for
the fractionalized PDW situation. The details for obtaining the same
is presented in Appendix~(\ref{App:Envelope}). To do this we performed
a few simplifications. First, we assume that the inter junction PP
to PH hopping amplitude is small compared to the intra-junction hoppings,
i.e. $t_{1}\ll t_{0}$. This is not necessary a priori but it simplifies
the following discussion. Experimentally, this requires hindering
the $t_{1}$ hopping by using a suitable barrier. Secondly, since
the envelope exists even in the long-time limit the transient response
can be safely ignored. Thirdly, the expressions of current is first
order perturbative solutions in $t_{0}$. We find that the expression
for the current in this limit becomes, 
\begin{align}
I_{J}\approx & \frac{n}{\hbar}\left[\sqrt{\kappa^{2}+t_{0}^{2}+2\kappa t_{0}\cos{\chi_{1}}}\right.\nonumber \\
 & \left.+\sqrt{\kappa^{2}+t_{0}^{2}+2\kappa t_{0}\cos{\chi_{2}}}\right]\sin\left(\theta_{0}\right).\label{eq:menv}
\end{align}
where $\chi_{1}=(\phi_{0}-\theta_{0})/2$ and $\chi_{2}=(\phi_{0}+\theta_{0})/2$.

The total envelope is shown by the red trace in Fig.~(\ref{Fig:fig4}a)
and Fig.~(\ref{Fig:fig4}b) which is the term inside the square bracket
in Eq.~(\ref{eq:menv}). The envelope term which captures the slower
oscillation is exhibited by the blue traces in Fig.~(\ref{Fig:fig4}).
We note that this is controlled by the $\chi_{1}$ term. Additionally,
the beat-like oscillations for the fractionalized PDW becomes better
resolved as the entanglement between the two order increases. Experimental
observation of such dependence will also signal the fractionalized
PDW in the pseudogapped phase of the underdoped cuprates.

\subsubsection{Extracting modulation wavevector}

\begin{figure}[t]
\includegraphics[width=8.5cm]{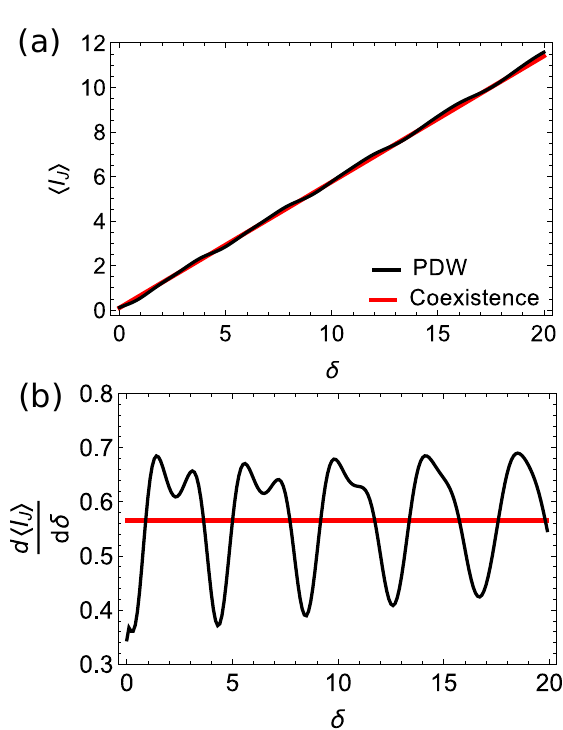} \caption{\textbf{(a)} Depicts the variation of average Josephson current with
the width of the insulating region, $\delta$. The current show a
linear increase for a simple coexistence of two orders, with the slope
proportional to the CDW modulation wavevector. However, for a fractionalized
PDW state, the Josephson current reveals a weak modulation with $\delta$.
\textbf{(b)} Shows the derivative of the current presented in the
(a). The PDW modulation wavevector $\mathbf{Q_{0}}$ controls the
oscillation of the Josephson current with $\delta$, albeit higher
harmonics of oscillations are also present.}
\label{Fig:fig5} 
\end{figure}

It is also possible to detect the PDW modulation wavevector by varying
the dephasing parameter $\delta$ between the two electrodes and by
investigating its effect on the average Josephson current. The initial
phases for the particle-hole pairs is given in Eq.~(\ref{eq:m11}),
$\theta^{\prime}=Q_{0}\delta$ and $\phi=Q_{0}\delta+\gamma$, where
$\gamma$ can be set to a constant. Similarly, the phase difference
for the particle-particle pairs, $\theta=C_{1}$, is also a constant
at $t=0$, which depends on the initial condition of the JJ setup.

We average the Josephson current, $\langle I_{J}\rangle$ using Eq.~(\ref{eq:mCurPDW})
over a time of 10 percent of the primary Josephson period $T$. For
the coexistence of order the $\langle I_{J}\rangle$ is presented
in Fig.~(\ref{Fig:fig5}a) in the red trace for $r_{0}=1.2$, $C_{1}=0$
and $Q_{0}=(1/4)2\pi/a_{0}$, where $a_{0}$ is the lattice spacing
set to unity. The expression is linearly increasing with the width
of the junction. The slope of the linear increase is proportional
to the modulation wavevector $Q_{0}$.

However, for the fractionalized PDW scenario, along with the linear
increase of the current with $\delta$, there is also weak oscillation
as shown in the black trace in Fig.~(\ref{Fig:fig5}a).
The modulation can be better identified by the derivative of $\langle I_{J}\rangle$
with respect to $\delta$. We depicted the same in Fig.~(\ref{Fig:fig5}b),
and it shows oscillations with the primary wavelength of $2\pi/Q_{0}$.
However, higher moments of oscillations make it challenging to determine
the magnitude of the PDW wavevector. (for details, see Appendix (\ref{Subsec:Epara}))
We also note that such Josephson junction is difficult to set up in
practice. In this setup, the width of the insulating region increases,
which should also modify the inter-junctions hopping for different
$\delta$. Moreover, fabricating JJ with varying sizes of the insulating
region is challenging.

\begin{figure}[t]
\includegraphics[width=9cm]{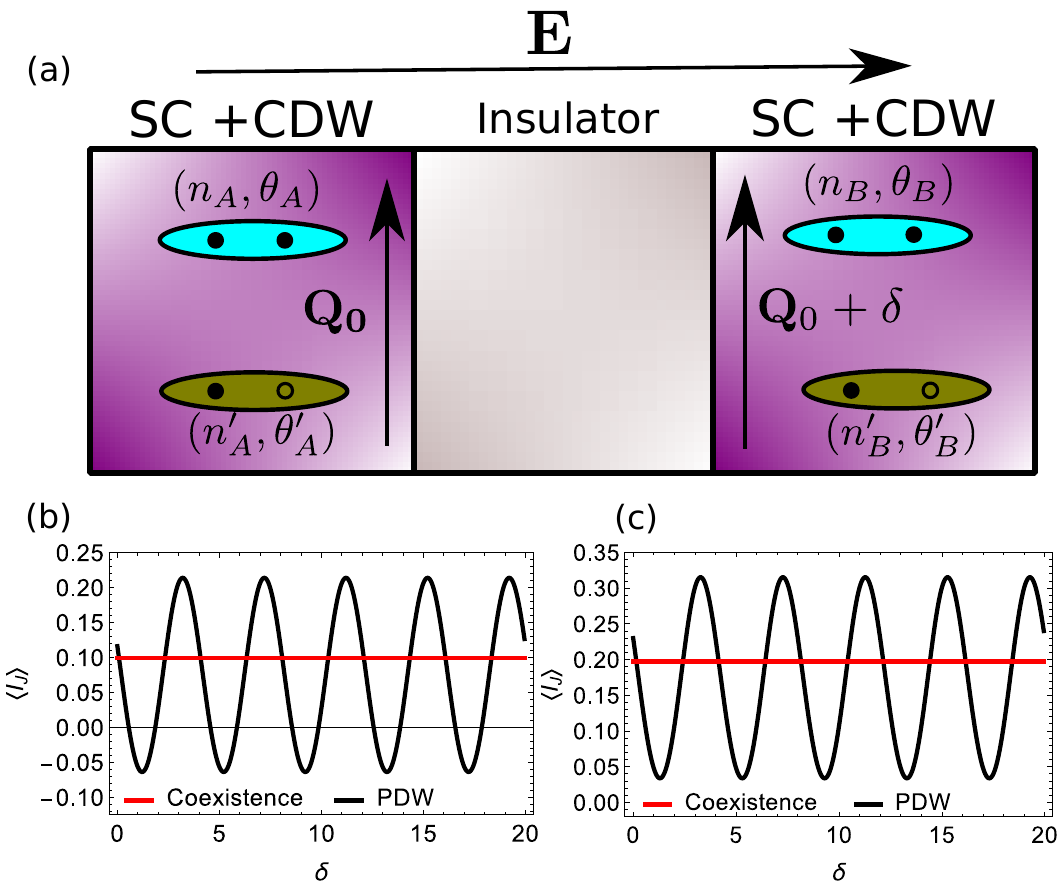} \caption{\textbf{(a)} Shows a Josephson junction setup in which the electric
field is perpendicular to the charge modulation wavevector. The CDW
modulations in the two terminals are dephased by a factor of $\delta$
here. \textbf{(b)} Depicts the evolution of the average Josephson
current with the dephasing parameter $\delta$. Here the $\langle I_{J}\rangle$
is time-averaged over ten percent of the primary Josephson period.
For the coexistence of orders, the average current with the dephasing
parameter is constant. However, for the fractionalized PDW scenario,
the Josephson current modulates with the modulation proportional to
the CDW wavevector $\mathbf{Q}_{0}$. \textbf{(c)} The same as (b)
but averaged over 20 percent of the primary Josephson period.}
\label{Fig:fig6} 
\end{figure}

Next, we discuss another complementary Josephson Junction setup better
suited in extracting the modulation wavevector $Q_{0}$. Fig.~(\ref{Fig:fig6}a)
shows a JJ setup where the modulation wavevector is perpendicular
to the electric field. Here $\delta$ denotes the phase-shift of the
CDW wavevector in the B electrode with respect to the A. In this situation, $V_{A}$
and $V_{B}$ vanishes and hence $r_{0}=\eta=0$. In Fig.~(\ref{Fig:fig6}b)
we plot the variation average Josephson current with $\delta$, time-averaged
over $0.1$ of the usual SC Josephson period. The details of the calculations
are presented in Appendix (\ref{Subsec:Eperp}). Here, the fractionalized
PDW scenario displays modulation controlled by $2\pi/Q_{0}$. In contrast,
the coexistence scenario gives a flat average current with varying
$\delta$. In Fig.~(\ref{Fig:fig6}c), we also establish that the
modulation of the $I_{J}$ with $\delta$ remains robust when time
averaging is done over twenty percent of the primary Josephson period.
Therefore, in this setup, it should be possible to extract the modulation
wavevector of the PDW.

\subsection{Inverse Josephson effect}

In the previous section, we used a constant DC-voltage $U$ across
the junction, leading to an AC-Josephson current. It is also possible
to apply a microwave AC-voltage to the junction, such that, $U(t)=U+\tilde{U}\cos{\omega t}$.
Here $U$ is the constant DC-Voltage. Note that we have used an insulating
barrier, so the normal current passing through the junction will vanish.
Solving for the Josephson current, for the simple coexistence of orders,
the Josephson current is given by 
\begin{align}
I_{J}=\frac{2\kappa n}{\hbar}\sum_{m=-\infty}^{\infty} & \left[(-1)^{m}\sin\left(C_{1}+\frac{2eU}{\hbar}t-m\omega t\right)\right.\nonumber \\
 & \left.\times\mathcal{J}_{m}\left(\frac{2e\tilde{U}}{\hbar\omega}\right)\right]+\frac{C_{2}r_{0}}{\pi\hbar}e^{-r_{0}t},\label{eq:IJE1}
\end{align}
where $m$ is an integer and $\mathcal{J}_{m}$ is the Bessel function
of the first kind of order $m$. The time average of this quantity
gives the DC current $I_{DC}$. The long time average of the oscillatory
term vanishes unless the frequency $\omega$ is some integral multiple
of the applied DC-Voltage $U$ in the units where we set $(2e)/\hbar=1$.
The transient current also fades away in the long time limit. The
DC-current in that scenario is given by, 
\begin{align}
\vert I_{DC}\vert\approx\frac{2\kappa n}{\hbar}\sum_{m=-\infty}^{\infty}\delta_{m\omega,U}(-1)^{m}\mathcal{J}_{m}\left(\frac{\tilde{U}}{\omega}\right)\sin{C_{1}}
\end{align}
which leads to sharp $\delta$ peaks at the integer multiples of the
AC-frequency $\omega$. These peaks are known as Shapiro spikes. We
have presented the details of Shapiro Spikes in Appendix~(\ref{App:SS}).

\begin{figure}[t]
\includegraphics[width=8.7cm]{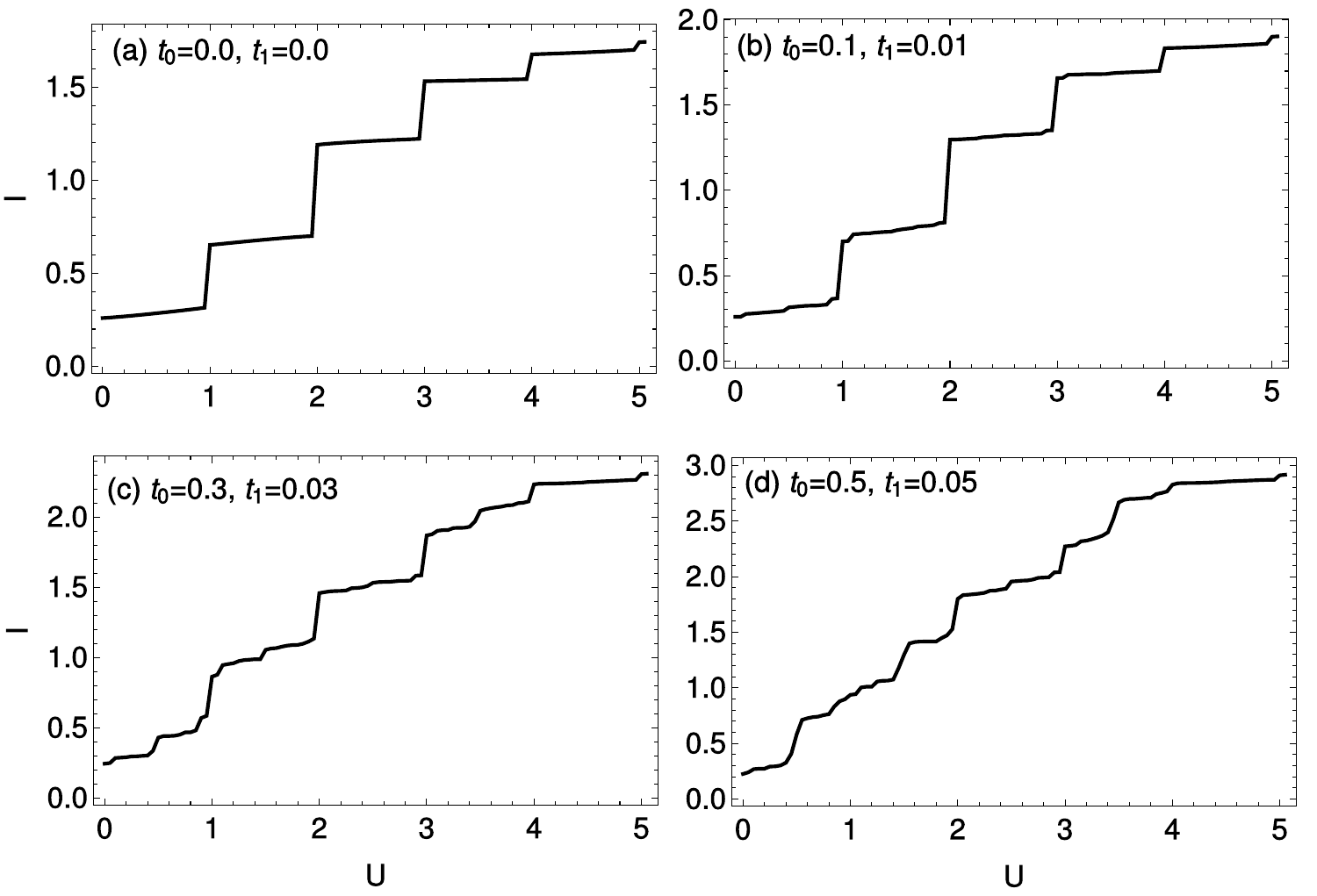} \caption{Demonstrates the current-voltage characteristics in the inverse Josephson
setup. In \textbf{(a)} we set $t_{0}=0,t_{1}=0$ the parameter corresponding
to the simple coexistence of CDW and SC orders. The sharp Shapiro
steps can be observed when the applied DC-voltage $U$ is equal to
the integer multiple of the frequency $\omega$. In \textbf{(b)-(d)}
we plot the same for finite $t_{0},t_{1}$ which corresponds to the
fractionalized PDW scenario. In \textbf{(b)} $t_{0},t_{1}$ is small,
and other steps start developing in between the two steps. In \textbf{(c)-(d)}
These additional steps become stronger as the entanglement between
the two orders is increased and some of the integer steps vanishes
completely. }
\label{Fig:fig7} 
\end{figure}

Usually, in experiments, the circuit is driven by current instead
of voltage. The nature of current versus voltage characteristics can
be qualitatively explained from the Shapiro spikes. For instance,
when the external driven current exceeds the strength of the Shapiro
spike at some voltage, the voltage increases abruptly with almost
zero slopes until the voltage reaches the next spike. In the subsequent
level, again, as the current increases to that of the spike strength,
the voltage remains stable. As the current exceeds the spike strength,
the voltage again shoots up until the next spike is reached. This
pattern keeps repeating itself, creating a step-like current-voltage
characteristic known as Shapiro steps.

We have solved Eqns.~(\ref{eq:m5-1}-\ref{eq:m8-1}) for an AC-voltage
of the form $U(t)=U+\tilde{U}\cos{\omega t}$ and plotted the DC current
as a function of the $U$ in the units where we set $(2e)/\hbar=1$
in Fig.~(\ref{Fig:fig7}). In all these figures we have used the
parameters same as in Fig.~(\ref{Fig:fig2}) with AC-frequency $\omega=1$
and AC voltage amplitude $\tilde{U}=3.0$, $C_{1}=\pi/2$. We have
presented the predicted current-driven nature of the current-voltage
characteristics in Fig.~(\ref{Fig:fig7}). For the coexistence case,
$t_{0},t_{1}=0$ and we find in Fig.~(\ref{Fig:fig7}a) the expected
sharp Shapiro steps at the integer multiple of $\omega$.

Next, we solve the Eqns.~(\ref{eq:m5-1}-\ref{eq:m8-1}) numerically
for an AC-voltage of the same form but a finite $t_{0},t_{1}$ and
track the evolution of the Shapiro steps. In Fig.~(\ref{Fig:fig7}b)
we present the results for $t_{0}=0.1,t_{1}=0.01$, i.e., a small
entanglement between the two orders leading to a weak PDW state. We
find that the Shapiro steps become broader as soon as the entanglement
between the two orders is turned on. New steps start to appear as
the overlap between the charge and SC order increases in Fig.~(\ref{Fig:fig7}c).
Finally, for a strong entanglement between the two orders, some steps
appear at different DC-voltage than the integer multiple of $\omega$.
Interestingly, in Fig.~(\ref{Fig:fig7}d), the first integer Shapiro
step at $U=1$ is thoroughly washed away as new step-like feature
forms at $U=1/2$. Therefore, our calculations suggest that additional
fractional Shapiro steps in the inverse Josephson junction setup will
strongly favor the PDW scenario. However, the conventional nature
of voltage-current characteristics will manifest for a competing order
scenario.

\section{Summary and conclusions}

\label{discus} We propose a Josephson junction setup that can distinguish
between two possible scenarios which can be observed in the enigmatic
pseudogap phase of the cuprates. We focused on the case of a fractionalized
PDW state which can turn into interconverted SC and CDW pairs and
compare this with the scenario where the SC and the CDW orders simply
coexist with each other. Our findings are that in the case of a fractionalized
PDW phase: 
\begin{itemize}
\item we observe a beat-like structure of AC-Josephson current when a constant
DC-Voltage is applied across the junction 
\item the additional frequencies for the AC Josephson current are at the
half-odd integer multiple of the normal Josephson frequency for a
large value of constant DC voltage. 
\item we can detect modulations of the Josephson current with a period proportional
to the CDW wavevector $\mathbf{Q}_{0}$ by varying the dephasing parameter
of the CDW modulation in two terminals. 
\item in the inverse Josephson setup, the induced DC-current has additional
steps other than the standard integer Shapiro steps. 
\end{itemize}
Identification of these signatures will strongly indicate the fractionalized
PDW scenario for the pseudogap phase. Recently signatures of PDW order
are seen in $\text{Bi}_{2}\text{Sr}_{2}\text{CaCu}_{2}\text{O}_{8+x}$
using a Josephson Scanning Tunnelling Microscopy setup~\cite{Hamidian16,PDW20_Nature}.
We, therefore, expect to observe these effects on such materials.
Moreover, the predictions presented here are not dependent on the
fine-tuning of material-dependent parameters.

We note that the Josephson Junction setup is shown in Fig.~(\ref{Fig:fig1})
and Fig.~(\ref{Fig:fig6}a) present just the two representative cases.
For instance, in Fig.~(\ref{Fig:fig1}), the applied electric field
is parallel to the modulation wavevector and in Fig.~(\ref{Fig:fig6}a),
the modulation wavevector is perpendicular to the applied field. However,
only short-ranged domains of unidirectional modulations are observed
in cuprates~\cite{comin2015broken}. Therefore, the samples in both
terminals can contain an admixture of unidirectional domains of modulated
orders. Importantly, our analysis shows that the presence of one scenario
cannot critically affect the other. Consequently, we expect to discern
a superposition of these two effects presented in the manuscript.

Recent studies have used Josephson scanning tunneling microscopy to
observe the modulation of the Josephson current~\cite{Hamidian16,PDW20_Nature}.
In such studies, both the tip and sample are in the underdoped regime
of the cuprates. The tips have been fabricated from the flakes of
the sample itself, leading to the possible detection of the pair density
wave states in cuprates. If the wavevectors of both the tip and sample
align perpendicular to the applied field, the situation of Fig.~(\ref{Fig:fig6}a)
will manifest. As the tip is moved parallelly over the sample, it
changes the dephasing parameter $\delta$. The Josephson current modulates
with the wavevector $\mathbf{Q}_{0}$ as a function of $\delta$~\cite{Hamidian16,PDW20_Nature},
very similar to our observation in Fig.~(\ref{Fig:fig6}).

Our calculations provide several predictions of the Josephson setups
to detect PDW states. Any experiments that can test these features
will be instrumental in differentiating between the simple coexistence
of orders and the proposed pair density wave scenario.

\section{Acknowledgement}

The authors thank Maxence Grandadam, J.C. S\'eamus Davis, and Yvan Sidis
for valuable discussions. This work has received financial support
from the ERC, under grant agreement AdG694651-CHAMPAGNE.

\appendix

\section{The Schr\"{o}dinger equations}

\label{AppEqns} In this section we provide complimentary details
on how to derive Eq.(\ref{eq:6}). We rewrite Eq.(\ref{eq:6}) as
\begin{align}
i\hbar\frac{\partial}{\partial t}\left(\sqrt{n_{A}}e^{i\theta_{A}}\right) & =\mu_{\Delta}\sqrt{n_{A}}e^{i\theta_{A}}+\kappa\sqrt{n_{B}}e^{i\theta_{B}}\label{eq:1-1}\\
 & +t_{0}\sqrt{n_{A}^{\prime}}e^{i\theta_{A}^{\prime}}+t_{1}\sqrt{n_{A}^{\prime}}e^{i\theta_{B}^{\prime}},\nonumber \\
i\hbar\frac{\partial}{\partial t}\left(\sqrt{n_{B}}e^{i\theta_{B}}\right) & =-\mu_{\Delta}\sqrt{n_{B}}e^{i\theta_{B}}+\kappa\sqrt{n_{A}}e^{i\theta_{A}}\label{eq:1-2}\\
 & +t_{0}\sqrt{n_{B}^{\prime}}e^{i\theta_{B}^{\prime}}+t_{1}\sqrt{n_{A}^{\prime}}e^{i\theta_{A}^{\prime}},\nonumber \\
i\hbar\frac{\partial}{\partial t}\left(\sqrt{n_{A}^{\prime}}e^{i\theta_{A}^{\prime}}\right) & =V_{A}\sqrt{n_{A}^{\prime}}e^{i\theta_{A}^{\prime}}+\kappa^{\prime}\sqrt{n_{B}^{\prime}}e^{i\theta_{B}^{\prime}}\label{eq:1-3}\\
 & +t_{0}\sqrt{n_{A}}e^{i\theta_{A}}+t_{1}\sqrt{n_{B}}e^{i\theta_{B}},\nonumber \\
i\hbar\frac{\partial}{\partial t}\left(\sqrt{n_{B}^{\prime}}e^{i\theta_{B}^{\prime}}\right) & =V_{B}\sqrt{n_{B}^{\prime}}e^{i\theta_{B}^{\prime}}+\kappa^{\prime}\sqrt{n_{A}^{\prime}}e^{i\theta_{A}^{\prime}}\label{eq:1-4}\\
 & +t_{0}\sqrt{n_{B}}e^{i\theta_{B}}+t_{1}\sqrt{n_{A}}e^{i\theta_{A}}.\nonumber 
\end{align}

Expanding Eq.(\ref{eq:1-1}) and taking the complex conjugate of it
yields 
\begin{align}
i\hbar\left(\dot{\sqrt{n_{A}}}+i\dot{\theta}_{A}\sqrt{n_{A}}\right)e^{i\theta_{A}} & =\mu_{\Delta}\sqrt{n_{A}}e^{i\theta_{A}}+\kappa\sqrt{n_{B}}e^{i\theta_{B}}\nonumber \\
 & +t_{0}\sqrt{n_{A}^{\prime}}e^{i\theta_{A}^{\prime}}+t_{1}\sqrt{n_{A}^{\prime}}e^{i\theta_{B}^{\prime}},\label{eq:2-1}\\
-i\hbar\left(\dot{\sqrt{n_{A}}}-i\dot{\theta}_{A}\sqrt{n_{A}}\right)e^{-i\theta_{A}} & =\mu_{\Delta}\sqrt{n_{A}}e^{-i\theta_{A}}+\kappa\sqrt{n_{B}}e^{-i\theta_{B}}\nonumber \\
 & +t_{0}\sqrt{n_{A}^{\prime}}e^{-i\theta_{A}^{\prime}}+t_{1}\sqrt{n_{A}^{\prime}}e^{-i\theta_{B}^{\prime}}.\label{eq:2-2}
\end{align}

Adding Eqs.(\ref{eq:2-1}) and (\ref{eq:2-2}) leads to 
\begin{align}
\frac{\partial n_{A}}{\partial t}=\frac{2}{\hbar} & \left[\kappa\sqrt{n_{A}n_{B}}\sin\theta+t_{0}\sqrt{n_{A}n_{A}^{\prime}}\sin(\theta_{A}^{\prime}-\theta_{A})\right.\nonumber \\
 & \left.+t_{1}\sqrt{n_{B}^{\prime}n_{A}}\sin(\theta_{B}^{\prime}-\theta_{A})\right],\label{eq:3}
\end{align}
where $\theta=\theta_{B}-\theta_{A}$. Next subtracting Eq.~(\ref{eq:2-1})
from Eq.~(\ref{eq:2-2}) leads to 
\begin{align}
\frac{\partial\theta_{A}}{\partial t}=-\frac{1}{\hbar} & \left[\mu_{\Delta}+\kappa\sqrt{\frac{n_{B}}{n_{A}}}\cos\theta+t_{0}\sqrt{\frac{n_{A}^{\prime}}{n_{A}}}\cos(\theta_{A}^{\prime}-\theta_{A})\right.\nonumber \\
 & \left.+t_{1}\sqrt{\frac{n_{B}^{\prime}}{n_{A}}}\cos(\theta_{B}^{\prime}-\theta_{A})\right],\label{eq:3-1}
\end{align}
Next repeating the same procedure for the Eq.~(\ref{eq:1-2}), we
obtain the corresponding equations for $n_{B}$ and $\theta_{B}$
as follows, 
\begin{align}
\frac{\partial n_{B}}{\partial t}=\frac{2}{\hbar} & \left[-\kappa\sqrt{n_{A}n_{B}}\sin\theta+t_{0}\sqrt{n_{B}n_{B}^{\prime}}\sin(\theta_{B}^{\prime}-\theta_{B})\right.\nonumber \\
 & \left.+t_{1}\sqrt{n_{A}^{\prime}n_{B}}\sin(\theta_{A}^{\prime}-\theta_{B})\right],\label{eq:4}\\
\frac{\partial\theta_{B}}{\partial t}=\frac{1}{\hbar} & \left[\mu_{\Delta}-\kappa\sqrt{\frac{n_{A}}{n_{B}}}\cos\theta-t_{0}\sqrt{\frac{n_{B}^{\prime}}{n_{B}}}\cos(\theta_{B}^{\prime}-\theta_{B})\right.\nonumber \\
 & \left.-t_{1}\sqrt{\frac{n_{A}^{\prime}}{n_{B}}}\cos(\theta_{A}^{\prime}-\theta_{B})\right].\label{eq:4-1}
\end{align}
We define $\bar{n}=n_{B}-n_{A}$, subtracting Eq.~(\ref{eq:3}) from
Eq.~(\ref{eq:4}), lead to 
\begin{align}
\frac{\partial\bar{n}}{\partial t} & =\frac{2}{\hbar}\left[-2\kappa\sqrt{n_{A}n_{B}}\sin\theta\nonumber\right.\\
 & \left.+t_{0}\left(\sqrt{n_{B}n_{B}^{\prime}}\sin(\theta_{B}^{\prime}-\theta_{B})-\sqrt{n_{A}n_{A}^{\prime}}\sin(\theta_{A}^{\prime}-\theta_{A})\right)\right.\nonumber \\
 & \left.+t_{1}\left(\sqrt{n_{A}^{\prime}n_{B}}\sin(\theta_{A}^{\prime}-\theta_{B})-\sqrt{n_{B}^{\prime}n_{A}}\sin(\theta_{B}^{\prime}-\theta_{A})\right)\right].\label{eq:5}
\end{align}
Approximating, that the density of of particle-particle pairs and
particle-hole pairs in both the terminals is similar in the steady
state, i.e., $n_{A}\approx n_{B}=n$ and $n_{A}^{\prime}\approx n_{B}^{\prime}=n^{\prime}$
and defining $\phi=\theta_{A}^{\prime}+\theta_{B}^{\prime}-\theta_{A}-\theta_{B}$,
we obtain 
\begin{align}
\frac{\partial\bar{n}}{\partial t}=\frac{4}{\hbar} & \left[-\kappa n\sin{\theta}+\sqrt{nn^{\prime}}\cos{\frac{\phi}{2}}\right.\nonumber \\
 & \left.\times\left(t_{0}\sin{\frac{\theta^{\prime}-\theta}{2}}-t_{1}\sin{\frac{\theta^{\prime}+\theta}{2}}\right)\right],\label{eq:5-1}
\end{align}
where $\theta^{\prime}=\theta_{B}^{\prime}-\theta_{A}^{\prime}$.
Following the same procedure and approximation we can obtain the differential
equation for $\theta$, which is given by 
\begin{align}
\frac{\partial\theta}{\partial t}=\frac{2}{\hbar} & \left[\mu_{\Delta}+\sqrt{\frac{n^{\prime}}{n}}\sin{\frac{\phi}{2}}\right.\nonumber \\
 & \left.\times\left(t_{0}\sin{\frac{\theta^{\prime}-\theta}{2}}-t_{1}\sin{\frac{\theta+\theta^{\prime}}{2}}\right)\right],\label{eq:6-1}
\end{align}

Repeating the procedure for Eqs. (\ref{eq:1-3}), (\ref{eq:1-4}),
we obtain the differential equation for the $\bar{n}^{\prime}$ and
$\theta^{\prime}$. 
\begin{align}
\frac{\partial\bar{n}^{\prime}}{\partial t}=\frac{4}{\hbar} & \left[-\kappa^{\prime}n^{\prime}\sin{\theta^{\prime}}+\sqrt{nn^{\prime}}\cos{\frac{\phi}{2}}\right.\nonumber \\
 & \left.\times\left(t_{0}\sin{\frac{\theta-\theta^{\prime}}{2}}-t_{1}\sin{\frac{\theta^{\prime}+\theta}{2}}\right)\right],\label{eq:7}\\
\frac{\partial\theta^{\prime}}{\partial t}=\frac{1}{\hbar} & \left[(V_{A}-V_{B})-2\sqrt{\frac{n^{\prime}}{n}}\sin{\frac{\phi}{2}}\right.\nonumber \\
 & \left.\times\left(t_{0}\sin{\frac{\theta-\theta^{\prime}}{2}}-t_{1}\sin{\frac{\theta^{\prime}+\theta}{2}}\right)\right],\label{eq:7-1}
\end{align}
Using all these forms for the $\theta$'s we can obtain the time evolution
equation for the $\phi$ 
\begin{align}
\frac{\partial\phi}{\partial t}= & -\frac{1}{\hbar}\left[(V_{A}+V_{B})+2\kappa\cos{\theta}+2\kappa^{\prime}\cos{\theta^{\prime}}\right.\nonumber \\
 & \left.-\frac{2(n-n^{\prime})}{\sqrt{nn^{\prime}}}\left(t_{0}\cos{\frac{\theta-\theta^{\prime}}{2}}+t_{1}\cos{\frac{\theta+\theta^{\prime}}{2}}\right)\right],\label{eq:8}
\end{align}
The fourth term on the RHS can be approximately taken to be small
when the particle-particle pairs and the particle-hole pairs are of
similar strength, i,e, $\delta n/n\ll1$ and thus we obtain 
\begin{align}
\frac{\partial\phi}{\partial t}\approx & -\frac{1}{\hbar}\left[(V_{A}+V_{B})+2\kappa\cos{\theta}+2\kappa^{\prime}\cos{\theta^{\prime}}\right].\label{eq:8-1}
\end{align}
We need to solve the five coupled differential Eqns. (\ref{eq:5-1}),
(\ref{eq:6-1}), (\ref{eq:7}), (\ref{eq:7-1}), and (\ref{eq:8-1}).

\section{Evaluation of Josephson current}

\label{AppJosephCurr} The Josephson current is obtained by the expression~\cite{JJbook,GrunerRMP}
\begin{align}
I_{J}=-\frac{1}{2}\frac{\partial\bar{n}}{\partial t}-\frac{1}{\pi}\frac{\partial\theta^{\prime}}{\partial t}.\label{eq:Jcur}
\end{align}
To find this, we need a solution to the equations presented in the
previous section. Using the forms for the difference of potential
between the two terminal due to charge density modulations as $V_{A}-V_{B}=-eE\bar{\rho}\theta^{\prime}\equiv-r_{0}\theta^{\prime}$,
with the average of the same $V_{A}+V_{B}=\eta$ set to constant.
The coupled differential equations can be written in a condensed form
as, 
\begin{align}
\frac{\partial\bar{n}}{\partial t} & =\frac{4}{\hbar}\left[-\kappa n\sin{\theta}+\sqrt{nn^{\prime}}\cos{\frac{\phi}{2}}f(t_{0},t_{1},\theta,\theta^{\prime})\right],\label{eq:11-1}\\
\frac{\partial\theta}{\partial t} & =\frac{2}{\hbar}\left[\mu_{\Delta}+\sqrt{\frac{n^{\prime}}{n}}\sin{\frac{\phi}{2}}f(t_{0},t_{1},\theta,\theta^{\prime})\right],\label{eq:11-2}\\
\frac{\partial\bar{n}^{\prime}}{\partial t} & =\frac{4}{\hbar}\left[-\kappa^{\prime}n^{\prime}\sin{\theta^{\prime}}+\sqrt{nn^{\prime}}\cos{\frac{\phi}{2}}g(t_{0},t_{1},\theta,\theta^{\prime})\right],\label{eq:11-3}\\
\frac{\partial\theta^{\prime}}{\partial t} & =\frac{1}{\hbar}\left[-r_{0}\theta^{\prime}-2\sqrt{\frac{n^{\prime}}{n}}\sin{\frac{\phi}{2}}g(t_{0},t_{1},\theta,\theta^{\prime})\right],\label{eq:11-4}\\
\frac{\partial\phi}{\partial t} & =-\frac{1}{\hbar}\left[\eta+2\kappa\cos{\theta}+2\kappa^{\prime}\cos{\theta^{\prime}}\right].\label{eq:11-5}
\end{align}
Here we have defined, 
\begin{align}
f(t_{0},t_{1},\theta,\theta^{\prime})=\left(t_{0}\sin{\frac{\theta^{\prime}-\theta}{2}}-t_{1}\sin{\frac{\theta^{\prime}+\theta}{2}}\right),\\
g(t_{0},t_{1},\theta,\theta^{\prime})=\left(t_{0}\sin{\frac{\theta-\theta^{\prime}}{2}}-t_{1}\sin{\frac{\theta+\theta^{\prime}}{2}}\right).
\end{align}
These equations can be solved using numerical means for any parameter,
and Josephson current can be evaluated using Eq.~(\ref{eq:Jcur}).
However, here we discuss an approach to calculate when the $t_{0}$
and $t_{1}$ are small parameters that can be incorporated perturbatively
in the expression. In this approach, we expand the solutions 
\begin{align}
\bar{n} & =\bar{n}_{0}+(t_{0}+t_{1})\bar{n}_{1}+(t_{0}+t_{1})^{2}\bar{n}_{2}+...\\
\theta^{\prime} & =\theta_{0}^{\prime}+(t_{0}+t_{1})\theta_{1}^{\prime}+(t_{0}+t_{1})^{2}\theta_{2}^{\prime}+...
\end{align}
and so on for other variables, and the subscript represents the perturbative
order of the solution. The zeroth-order solution is readily obtained
by putting $t_{0}=0$ and $t_{1}=0$ in Eq.~(\ref{eq:11-1}) to Eq.~(\ref{eq:11-5}).
The relevant equations becomes, 
\begin{align}
\frac{\partial\bar{n}_{0}}{\partial t} & =-\frac{4\kappa n}{\hbar}\sin{\theta_{0}},\label{eq:12-1}\\
\frac{\partial\theta_{0}}{\partial t} & =\frac{2\mu_{\Delta}}{\hbar},\label{eq:12-2}\\
\frac{\partial\theta_{0}^{\prime}}{\partial t} & =-\frac{r_{0}}{\hbar}\theta_{0}^{\prime},\label{eq:12-3}\\
\frac{\partial\phi_{0}}{\partial t} & =-\frac{1}{\hbar}\left[\eta+2\kappa\cos{\theta_{0}}+2\kappa^{\prime}\cos{\theta_{0}^{\prime}}\right].\label{eq:12-4}
\end{align}
The solution for these equations are readily obtained and these are
given by, 
\begin{align}
\theta_{0} & =\frac{2\mu_{\Delta}}{\hbar}t+C_{1},\label{eq:theta0}\\
\theta_{0}^{\prime} & =\frac{C_{2}}{\hbar}e^{-r_{0}t},\label{eq:theta0p}
\end{align}
and the zeroth order Josephson current becomes, 
\begin{align}
I_{J}^{(0)}=\frac{2\kappa n}{\hbar}\sin\left(\frac{2\mu_{\Delta}}{\hbar}t+C_{1}\right)+\frac{C_{2}r_{0}}{\pi\hbar}e^{-r_{0}t},\label{eq:cur_coex}
\end{align}
where $C_{1}$ and $C_{2}$ are the constants of integration. Notice
that the first term is the usual Josephson current for a superconducting
junction, whereas the second term is a transient current due to the
presence of charge orders. When the entanglement between the orders
are small, i.e. for a simple coexistence of order Eq.~(\ref{eq:cur_coex})
gives the exact form of the current.

Next we focus on obtaining the first order correction to this current.
To this end, we need the zeroth order expression for $\phi_{0}$,
which can be obtained by using $\theta_{0}$ and $\theta_{0}^{\prime}$
in Eq.~(\ref{eq:12-4}), 
\begin{align}
\phi_{0}=-\frac{\eta}{\hbar}t+C_{3}+\frac{2\kappa^{\prime}}{r_{0}\hbar}\text{Ci}\left[C_{2}e^{-r_{0}t}\right]-\frac{\kappa}{\mu_{\Delta}}\sin\left(\frac{2\mu_{\Delta}}{\hbar}t+C_{1}\right),\label{eq:phi0}
\end{align}
where again $C_{3}$ is a constant of integration and $\text{Ci}[x]$
is the cosine integral function. The first order, equations can now
be evaluated by taking the derivative of Eq.~(\ref{eq:11-1}) and
Eq.~(\ref{eq:11-4}) with respect to $t_{0}$ and $t_{1}$ individually
and subsequently setting these small parameter to zero. Therefore
the first-order correction for the terms relevant for the Josephson
current thus becomes 
\begin{align}
\frac{\partial\bar{n}_{1}}{\partial t} & =\frac{4\sqrt{nn^{\prime}}}{\hbar}\cos{\frac{\phi_{0}}{2}}\left(t_{0}\sin{\frac{\theta_{0}^{\prime}-\theta_{0}}{2}}-t_{1}\sin{\frac{\theta_{0}^{\prime}+\theta_{0}}{2}}\right),\label{eq:13-1}\\
\frac{\partial\theta_{1}^{\prime}}{\partial t} & =-\frac{2}{\hbar}\sqrt{\frac{n^{\prime}}{n}}\sin{\frac{\phi_{0}}{2}}\left(t_{0}\sin{\frac{\theta-\theta_{0}^{\prime}}{2}}-t_{1}\sin{\frac{\theta_{0}+\theta_{0}^{\prime}}{2}}\right).\label{eq:13-2}
\end{align}
Therefore up to the first order the Josephson current is given by,
\begin{align}
I_{J}= & \frac{2\kappa n}{\hbar}\sin\left(\frac{2\mu_{\Delta}}{\hbar}t+C_{1}\right)+\frac{C_{2}r_{0}}{\pi\hbar}e^{-r_{0}t}\nonumber \\
 & -\frac{2\sqrt{nn^{\prime}}}{\hbar}\cos{\frac{\phi_{0}}{2}}\left(t_{0}\sin{\frac{\theta_{0}^{\prime}-\theta_{0}}{2}}-t_{1}\sin{\frac{\theta_{0}^{\prime}+\theta_{0}}{2}}\right)\nonumber \\
 & +\frac{2}{\pi\hbar}\sqrt{\frac{n^{\prime}}{n}}\sin{\frac{\phi_{0}}{2}}\left(t_{0}\sin{\frac{\theta_{0}-\theta_{0}^{\prime}}{2}}-t_{1}\sin{\frac{\theta_{0}+\theta_{0}^{\prime}}{2}}\right),\label{eq:CurPDW}
\end{align}
where we can use the time evolution of $\theta_{0}$, $\theta_{0}^{\prime}$,
and $\phi_{0}$ from Eq.~(\ref{eq:theta0}), Eq.~(\ref{eq:theta0p})
and Eq.~(\ref{eq:phi0}) respectively. In the main text, this form
is compared favorably with the exact current evaluated by solving
the equations numerically.

\section{Extracting the frequencies of the AC Josephson current}

\label{App:Freq} The Josephson current for a fractionalized PDW state
shows a beat like structure. This suggests multiple frequencies are
contributing to the AC Josephson current. This section provides the
details to obtain the Josephson current frequencies. To do so, we
need to perform a Fourier transform of the AC Josephson current to
the frequency domain.

We simplify the Eq.~(\ref{eq:CurPDW}) by assuming that the inter
junction PP to PH hopping amplitude is small compared to the intra-junction
hoppings, i.e., $t_{1}\ll t_{0}$. In an experimental scenario, this
requires using a barrier to decay the PH hoppings across the junction.
However, for a finite but small $t_{1}$, will not create any qualitative
difference to the discussion below. Also, in a long time limit, the
transient current regime from the CDW vanishes, i.e., $\theta_{0}^{\prime}\rightarrow0$.
Hence the current reduces to 
\begin{align}
I_{J}\approx & \frac{2\kappa n}{\hbar}\sin{\theta_{0}}+\frac{2t_{0}\sqrt{nn^{\prime}}}{\hbar}\cos{\frac{\phi_{0}}{2}}\sin{\frac{\theta_{0}}{2}}\nonumber \\
 & -\frac{2t_{0}}{\pi\hbar}\sqrt{\frac{n^{\prime}}{n}}\sin{\frac{\phi_{0}}{2}}\sin{\frac{\theta_{0}}{2}},\label{eq:AprxPDW1}
\end{align}

Furthermore, since the $n\approx n^{\prime}$, the second term dominates
over the third. The expression for the current further simplifies
to 
\begin{align}
I_{J}\approx & \frac{2\kappa n}{\hbar}\sin{\theta_{0}}+\frac{2t_{0}\sqrt{nn^{\prime}}}{\hbar}\cos{\frac{\phi_{0}}{2}}\sin{\frac{\theta_{0}}{2}}.\label{eq:AprxPDW2}
\end{align}
Using trignometric identities, the second term becomes, 
\begin{align}
I_{2}\approx & \frac{t_{0}\sqrt{nn^{\prime}}}{\hbar}\left[\sin\left(\frac{\theta_{0}+\phi_{0}}{2}\right)+\sin\left(\frac{\theta_{0}-\phi_{0}}{2}\right)\right].\label{eq:AprxPDW3}
\end{align}

The $\theta_{0}$ and $\phi_{0}$ is given by Eq.~(\ref{eq:theta0})
and Eq.~(\ref{eq:phi0}) respectively. The constant terms do not
contribute to the frequency of the AC Josephson frequency, and hence
we can ignore them for the following analysis. Next we make a series
expansion for $Ci[...]$ function and neglecting the constant and
higer order terms, we obtain 
\begin{align}
I_{2}\approx & \frac{t_{0}\sqrt{nn^{\prime}}}{\hbar}\left[\sin\left(a_{+}+b\sin{\xi t}\right)+\sin\left(a_{-}-b\sin{\xi t}\right)\right],\label{eq:AprxPDW4}
\end{align}
where we have defined 
\begin{align}
a_{\pm} & =\frac{1}{2}\left(\frac{2\mu_{\Delta}}{\hbar}\pm\frac{\eta}{\hbar}\pm\frac{2\kappa^{\prime}}{\hbar}\right)\\
b & =\frac{\kappa}{2\mu_{\Delta}}\\
\xi & =\frac{2\mu_{\Delta}}{\hbar}
\end{align}
Next, we expand 
\begin{align}
\cos(x\sin\theta)=\mathcal{J}_{0}(x)+2\sum_{p=1}^{\infty}\mathcal{J}_{2p}(x)\cos(2p\theta),\\
\sin(x\sin\theta)=2\sum_{p=1}^{\infty}\mathcal{J}_{2p+1}(x)\sin((2p+1)\theta),
\end{align}
Where $\mathcal{J}_{\nu}(x)$ is the Bessel function of first kind
of the $\nu$-th order. Neglecting the higher order terms the current
in Eq.~(\ref{eq:AprxPDW1}) becomes 
\begin{align}
I_{J}\approx & \frac{t_{0}\sqrt{nn^{\prime}}}{\hbar}\left[\mathcal{J}_{0}(b)\left\lbrace \sin a_{+}t+\sin a_{-}t\right\rbrace +\right.\nonumber \\
 & \left.\mathcal{J}_{1}(b)\left\lbrace \sin\left((a_{+}+\xi)t\right)-\sin\left((a_{-}+\xi)t\right)\right.\right.\nonumber \\
 & \left.\left.+\sin a_{-}t-\sin a_{+}t\right\rbrace \right]+\frac{2\kappa n}{\hbar}\sin{\xi t}
\end{align}
Since, all the terms are directly proportional to $t$, one can easily
read off the frequencies for the AC current, by performing a fourier
transform. This is given by 
\begin{align}
\tilde{I}_{J}(\omega)= & \frac{\kappa n}{\hbar}\delta({\omega-\xi})+\frac{t_{0}\sqrt{nn^{\prime}}}{2\hbar}\left[(\mathcal{J}_{0}(b)+\mathcal{J}_{1}(b))\delta(\omega-a_{-})\right.\nonumber \\
 & \left.+\mathcal{J}_{1}(b)\delta(\omega-a_{+}-\xi)-\mathcal{J}_{1}(b)\delta(\omega-a_{-}-\xi)\right.\nonumber \\
 & \left.+(\mathcal{J}_{0}(b)-\mathcal{J}_{1}(b))\delta(\omega-a_{-})\right]\label{eq:freq}
\end{align}
The primary frequency $\xi$ is the usual AC Josephson frequency.
The other frequencies are the additional ones originating due to the
entanglement of the two orders. The ratio of the primary to the few
additional frequencies is given by,

\begin{align}
\frac{\omega_{1}}{\omega_{0}}=\frac{1}{2}-\left(\frac{\eta+2\kappa^{\prime}}{4eU}\right),\\
\frac{\omega_{4}}{\omega_{0}}=\frac{1}{2}+\left(\frac{\eta+2\kappa^{\prime}}{4eU}\right),\\
\frac{\omega_{2}}{\omega_{0}}=\frac{3}{2}+\left(\frac{\eta+2\kappa^{\prime}}{4eU}\right),\\
\frac{\omega_{3}}{\omega_{0}}=\frac{3}{2}-\left(\frac{\eta+2\kappa^{\prime}}{4eU}\right),
\end{align}
where $\omega_{i-1}$ are the $i$-th delta function peak of Eq.~(\ref{eq:freq}).
For large potential difference between the two junctions i.e., $U\gg\eta+2\kappa^{\prime}$,
the second term of all the ratios vanish. Moreover, the ratio between
the primary and additional frequencies will occur at half-odd integers,
i.e., $1/2,3/2,5/2,...$. However, the peak strength will diminish
for the higher order term as it is determined by the higher order
Bessel function. Studying such frequency dependence of the AC Josephson
current will give an indication of the fractionalized PDW order.

\section{Extracting the envelope}

\label{App:Envelope} The Josephson current for a PDW state or fractionalized
PDW state shows a beat like structure. Such a beat-like form can be
distinguished in experiments establishing an entanglement between
the superconducting and the charge orders. To provide a detailed description,
it becomes necessary to determine the parameters that control such
a current envelope. In this section, we provide the details of the
envelope of the Josephson current. We start with Eq.(\ref{eq:AprxPDW2})
Using trignometric identities, we obtain 
\begin{align}
I_{J}= & \frac{\kappa n}{\hbar}\sin{\theta_{0}}+\frac{t_{0}\sqrt{nn^{\prime}}}{\hbar}\sin\left(\theta_{0}+\frac{\phi_{0}-\theta_{0}}{2}\right)\nonumber \\
 & +\frac{\kappa n}{\hbar}\sin{\theta_{0}}+\frac{t_{0}\sqrt{nn^{\prime}}}{\hbar}\sin\left(\theta_{0}-\frac{\phi_{0}+\theta_{0}}{2}\right).\label{eq:AprxPDW3_3}
\end{align}
Before adding the sine waves we define, $(\phi_{0}-\theta_{0})/2=\chi_{1}$
and $(\phi_{0}+\theta_{0})/2=\chi_{2}$. Therefore, using $n\approx n^{\prime}$,
the expression current becomes, 
\begin{align}
I_{J}= & \frac{n}{\hbar}\left[\sqrt{\kappa^{2}+t_{0}^{2}+2\kappa t_{0}\cos{\chi_{1}}}\sin\left(\theta_{0}+\xi_{1}\right)\right.\nonumber \\
 & \left.+\sqrt{\kappa^{2}+t_{0}^{2}+2\kappa t_{0}\cos{\chi_{2}}}\sin\left(\theta_{0}-\xi_{2}\right)\right].
\end{align}
Here the phase angles $\xi_{i}$ are given by 
\begin{align}
\xi_{i}=\sin^{-1}\left(\frac{t_{0}\sin{\chi_{i}}}{\sqrt{\kappa^{2}+t_{0}^{2}+2\kappa t_{0}\cos{\chi_{i}}}}\right),
\end{align}
where $i=1,2$. In the perturbative limit, $t_{0}$ is the small parameter
in our calculations, the phase shifts can be assumed to be small,
such that $\xi_{i}=0$. Therefore the current becomes 
\begin{align}
I_{J}= & \frac{n}{\hbar}\sum_{i=1}^{2}\left[\sqrt{\kappa^{2}+t_{0}^{2}+2\kappa t_{0}\cos{\chi_{i}}}\right]\sin\left(\theta_{0}\right).\label{eq:ApprxCurrmod}
\end{align}
Here the amplitude of the envelope is controlled by two phases. The
faster oscillation is governed by $\chi_{2}=(\phi_{0}+\theta_{0})/2$
and the slower one by $\chi_{1}=(\phi_{0}-\theta_{0})/2$. The total
envelope is the sum of the square root term in Eq.~(\ref{eq:ApprxCurrmod}),
and therefore contains further oscillations given by $\chi_{1}$ and
$\chi_{2}$.

\section{Josephson current with dephasing parameter}

\label{JosephsonDephase} 

\subsection{Electric field parallel to $\mathbf{Q_{0}}$}

\label{Subsec:Epara} This appendix discusses the procedure to obtain
the average Josephson current with the dephasing parameter. Initially
at $t=0$, the phase of the CDW order is given by $\theta^{\prime}(0)=Q_{0}\delta=C_{2}$.
Similarly, one can set $\phi(0)=Q_{0}\delta+\gamma$, where $\gamma$
is a constant. The zeroth order solution for $\phi$, thus becomes
\begin{align}
\phi_{0}=-\frac{\eta}{\hbar}t+(Q_{0}\delta+\gamma)+\frac{2\kappa^{\prime}}{r_{0}\hbar}\text{Ci}\left[Q_{0}\delta e^{-r_{0}t}\right]\nonumber \\
-\frac{\kappa}{\mu_{\Delta}}\sin\left(\frac{2\mu_{\Delta}}{\hbar}t+C_{1}\right).\label{eq:mCurPDW0}
\end{align}
Putting this in Eq.~(\ref{eq:mCurPDW}), we obtain the $I_{J}(t)$,
which we integrate numerically over a fixed time to get the average
$\langle I_{J}\rangle$. Notice, the cosine-integral dependence of
$\delta$ in $\phi_{0}$ generates a higher harmonics of oscillation
of $\langle I_{J}\rangle$ with $\delta$. Thus, it becomes difficult
to extract the PDW modulation wavevector in this setup.

\subsection{Electric field perpendicular to $\mathbf{Q_{0}}$}

\label{Subsec:Eperp} No such complications arise when the JJ is set
up is such that the wavevector is perpendicular to the electric field.
In this case, although $r_{0}=\eta=0$ and the transient current cannot
survive, yet initial $\theta_{0}^{\prime}(0)=Q_{0}\delta$. Similarly,
one can set $\phi(0)=Q_{0}\delta+\gamma$, where $\gamma$ is a constant.
Here $\delta$ denotes the phase difference between the CDW wavevector
in the two terminals. Consequently, the zeroth-order solutions become
\begin{align}
\theta_{0} & =\frac{2\mu_{\Delta}}{\hbar}t+C_{1},\label{eq:Eperpmtheta0}\\
\theta_{0}^{\prime} & =Q_{0}\delta,\label{eq:Eperpmtheta0p}\\
\phi_{0} & =(Q_{0}\delta+\gamma)-\frac{2\kappa^{\prime}}{\hbar}\cos\left(Q_{0}\delta\right)t-\frac{\kappa}{\mu_{\Delta}}\sin\left(\frac{2\mu_{\Delta}}{\hbar}t+C_{1}\right).\label{eq:Eperpmphi0}
\end{align}
Putting these expressions in the Eq.~(\ref{eq:mCurPDW}), we can
evaluate average current, which shows clear modulations with $\delta$.

\section{Current due to CDW phase}

\begin{figure}[t]
\includegraphics[width=7cm]{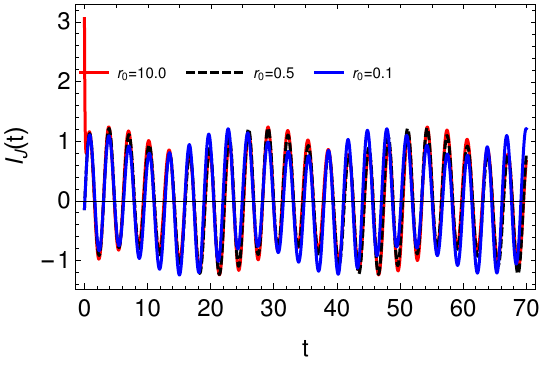} \caption{Shows the AC Josephson current for different values of $r_{0}$. The
beat-like structure for the AC current remains robust for all the
values of $r_{0}$. Only the intial transient nature of the current
vanishes for smaller $r_{0}$.}
\label{Fig:FigApp3} 
\end{figure}

The CDW phase can be strongly pinned by the boundary of the junctions
and the disorder of the samples. However, here we consider the general
possibility for the current arising due to the phase of the CDW order.
We note that the term ${V_{A}-V_{B}=-(eEn^{\prime}/Q_{0})\theta^{\prime}\equiv-r_{0}\theta^{\prime}}$,
generates only a transient current response at short times. The parameter
$r_{0}$ characterizes the strength of the phase current from the
CDW. The long-time behavior of the AC Josephson current remains invariant
when we vary this term. In Fig.~(\ref{Fig:FigApp3}), we have shown
the evolution of AC Josephson current for the PDW scenario. The parameters
used here are the same as those presented in Fig.~(2b) of the main
text. As we increase the $r_{0}$, the AC response changes for $t\rightarrow0$,
whereas it remains invariant at long times. The frequency of the oscillation
is independent of $r_{0}$ and thus remains the same. The amplitude
of the oscillation changes weakly when $r_{0}<1$. Therefore, the
results presented in the manuscript remain robust while we change
this parameter.

\section{Josephson current with complex hoppings}

\label{App:Complex} Here we consider the complex hoppings between
the particle-particle to particle-hole hoppings. Since this term is
connected by the PDW order, it can be complex in general. We confirm
here whether the imaginary part of these hoppings consideribly modifies
our results.

The set of Schr\"{o}dinger's equations write, with $\overline{\psi}=\left(\begin{array}{cccc}
\sqrt{n_{A}}e^{i\theta_{A}}, & \sqrt{n_{B}}e^{i\theta_{B}}, & \sqrt{n_{A}^{\prime}}e^{i\theta_{A}^{\prime}}, & \sqrt{n_{B}^{\prime}}e^{i\theta_{B}^{\prime}}\end{array}\right)$ 
\begin{align}
i\hbar\frac{\partial}{\partial t}\psi & =\left(\begin{array}{cccc}
\mu_{\Delta} & \kappa & t_{0} & t_{1}\\
\kappa & -\mu_{\Delta} & t_{1} & t_{0}\\
t_{0}^{*} & t_{1}^{*} & V_{A} & \kappa^{\prime}\\
t_{1}^{*} & t_{0}^{*} & \kappa^{\prime} & V_{B}
\end{array}\right)\psi,\label{eq:App_eq1}
\end{align}

By following the pocedure similar to Appendix.~(\ref{AppEqns}),
we arrive at the set of Schr\"{o}dinger equations 
\begin{widetext}
\begin{align}
\frac{\partial\bar{n}}{\partial t} & =\frac{4}{\hbar}\left[-\kappa n\sin\theta+\sqrt{nn^{\prime}}\left\lbrace \cos\frac{\phi}{2}\left(\text{Re }t_{0}\sin{\frac{\theta^{\prime}-\theta}{2}}-\text{Re }t_{1}\sin{\frac{\theta^{\prime}+\theta}{2}}\right)-\sin\frac{\phi}{2}\left(\text{Im }t_{0}\sin{\frac{\theta^{\prime}-\theta}{2}}-\text{Im }t_{1}\sin{\frac{\theta^{\prime}+\theta}{2}}\right)\right\rbrace \right],\\
\frac{\partial\theta}{\partial t} & =\frac{2}{\hbar}\left[\mu_{\Delta}+\sqrt{\frac{n^{\prime}}{n}}\left\lbrace \sin\frac{\phi}{2}\left(\text{Re }t_{0}\sin{\frac{\theta^{\prime}-\theta}{2}}-\text{Re }t_{1}\sin{\frac{\theta^{\prime}+\theta}{2}}\right)-\cos\frac{\phi}{2}\left(\text{Im }t_{0}\sin{\frac{\theta^{\prime}-\theta}{2}}-\text{Im }t_{1}\sin{\frac{\theta^{\prime}+\theta}{2}}\right)\right\rbrace \right],\\
\frac{\partial\bar{n}^{\prime}}{\partial t} & =\frac{4}{\hbar}\left[-\kappa^{\prime}n^{\prime}\sin\theta^{\prime}+\sqrt{nn^{\prime}}\left\lbrace \cos\frac{\phi}{2}\left(\text{Re }t_{0}\sin{\frac{\theta-\theta^{\prime}}{2}}-\text{Re }t_{1}\sin{\frac{\theta^{\prime}+\theta}{2}}\right)-\sin\frac{\phi}{2}\left(\text{Im }t_{0}\sin{\frac{\theta-\theta^{\prime}}{2}}-\text{Im }t_{1}\sin{\frac{\theta^{\prime}+\theta}{2}}\right)\right\rbrace \right],\\
\frac{\partial\theta^{\prime}}{\partial t} & =\frac{1}{\hbar}\left[V_{A}-V_{B}-2\sqrt{\frac{n}{n^{\prime}}}\left\lbrace \sin\frac{\phi}{2}\left(\text{Re }t_{0}\sin{\frac{\theta-\theta^{\prime}}{2}}-\text{Re }t_{1}\sin{\frac{\theta^{\prime}+\theta}{2}}\right)-\cos\frac{\phi}{2}\left(\text{Im }t_{0}\sin{\frac{\theta-\theta^{\prime}}{2}}-\text{Im }t_{1}\sin{\frac{\theta^{\prime}+\theta}{2}}\right)\right\rbrace \right],\\
\frac{\partial\phi}{\partial t} & =-\frac{1}{\hbar}\left[\eta+2\kappa\cos{\theta}+2\kappa^{\prime}\cos{\theta^{\prime}}\right].
\end{align}
\end{widetext}

\begin{figure}[t]
\includegraphics[width=7cm]{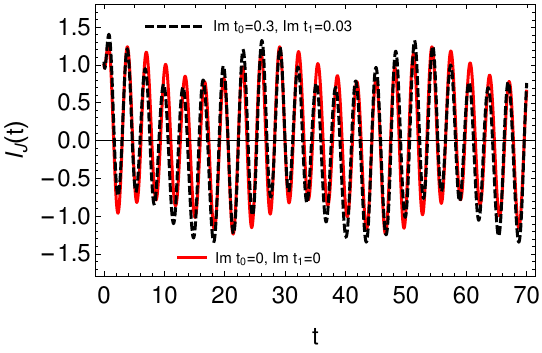} \caption{AC Josephson current for two scenarios. In both the $\text{Re }t_{0}=0.3$
and $\text{Re }t_{1}=0.03$. Moreover, all other parameters are the
same as we used in Fig.~(\ref{Fig:fig2}b). The dashed lines manifest
the effects of the imaginary part of the $t_{0}$, and $t_{1}$ is
finite. Although the amplitude of the current changes slightly as
we increase the imaginary part of the hoppings, the oscillation frequencies
remain the same.}
\label{Fig:FigApp1} 
\end{figure}

We solve the equations numerically and derive the Josephson current
using Eq.(\ref{eq:m9}). The results are presented in Fig.~(\ref{Fig:FigApp1}).

It shows the Josephson current as the imaginary part of the hoppings
are increased. Although the amplitude of the current changes with
the increasing imaginary part of the hoppings, the frequency remains
the same. The oscillation frequencies in Eq.~(\ref{eq:freq1}-\ref{eq:freq4})
are independent of the $t_{0}$ and $t_{1}$. Consequently, our other
results presented in the main paper remains robust if we consider
these hoppings to be complex. The results are also expected to hold
when even if we consider different values for $t_{0}$ in two terminals.

\section{Shapiro spikes}

\label{App:SS} 
\begin{figure}[ht]
\includegraphics[width=8.75cm]{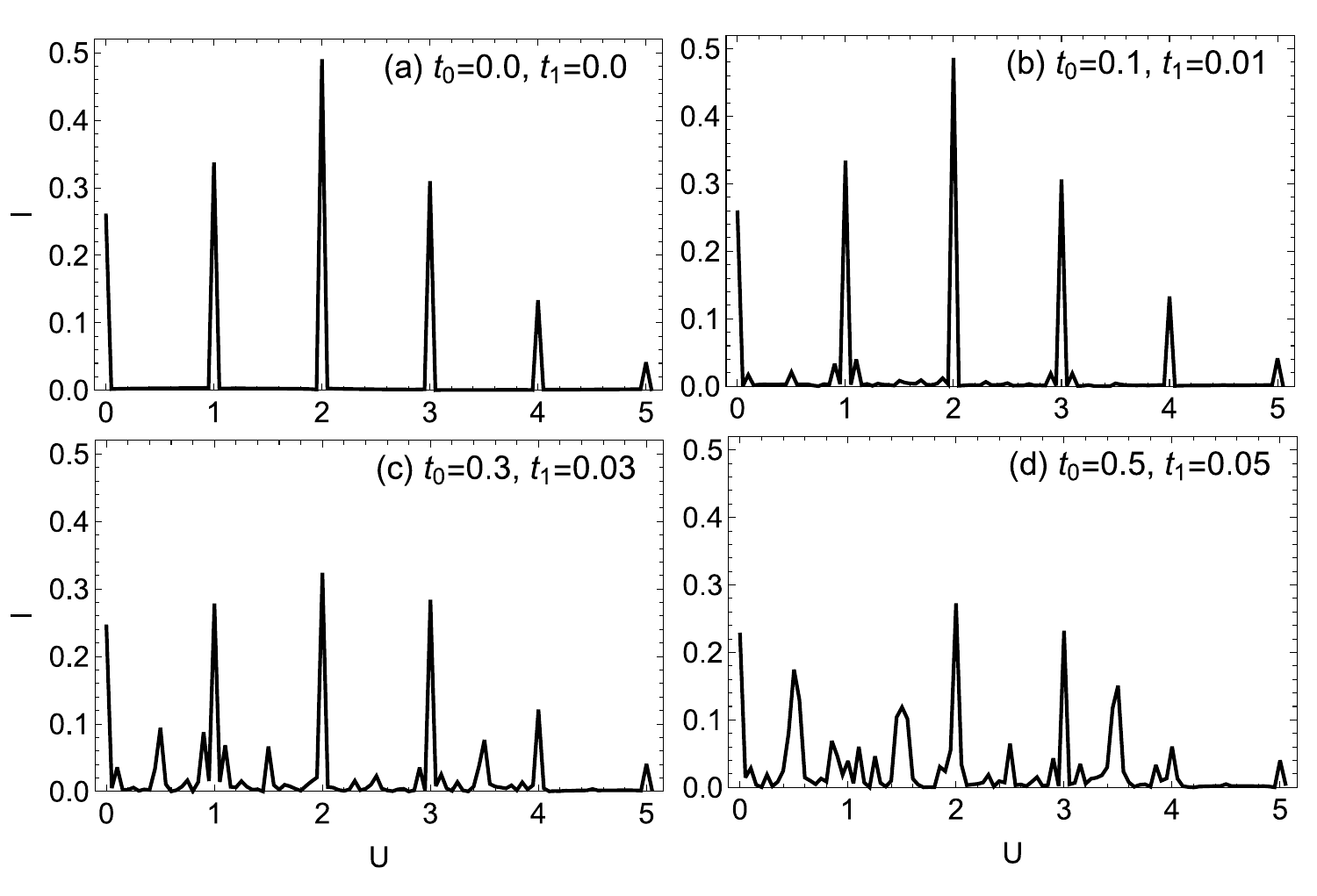} \caption{Demonstrates the total DC-current with the applied DC voltage $U$
across the junctions in the inverse Josephson setup. In \textbf{(a)}
we set $t_{0}=0,t_{1}=0$ the parameter corresponding to the simple
coexistence of CDW and SC orders. The sharp Shapiro spikes can be
observed when the applied DC-voltage $U$ is equal to the integer
multiple of the applied frequency $\omega$. In \textbf{(b)-(d)} we
plot the same for finite $t_{0},t_{1}$ which corresponds to the fractionalized
PDW scenario. In \textbf{(b)} $t_{0},t_{1}$ is small, and other peaks
start developing in between the two Shapiro spikes. In \textbf{(c)-(d)}
These additional peaks become stronger as the entanglement between
the two orders is increased. }
\label{Fig:FigApp2} 
\end{figure}

The main paper shows the qualitative voltage-current characteristics
in a current-driven Josephson circuit, similar to the experimental
situation. Here we present the Shapiro spikes for the voltage-driven
Josephson systems for completeness. In this setup, the current shows
sharp $\delta$ peaks at the integer multiples of the AC-frequency
$\omega$ which is set to unity. These peaks are known as Shapiro
spikes.

For the coexistence of orders $t_{0},t_{1}=0$ and we find in Fig.~(\ref{Fig:FigApp2}a)
the expected sharp Shapiro spikes at the integer multiple of $\omega$.
We also solve the Eqns.~(\ref{eq:m5-1}-\ref{eq:m8-1}) numerically
for an AC-voltage of the same form but a finite $t_{0},t_{1}$ and
track the evolution of the Shapiro spikes. In Fig.~(\ref{Fig:FigApp2}b)
we present the results for a weak entanglement between the two orders.
We see that multiple weak peaks emerge as soon as the entanglement
between the two orders increases. Such extra peaks get stronger as
the overlap between the charge and SC order becomes large in Fig.~(\ref{Fig:FigApp2}c).
Finally, for a strong fractionalized PDW state, the Shapiro spikes
appear at different DC-voltage than the integer multiple of $\omega$.
For instance, a spike appears at $U=1/2$ in Fig.~(\ref{Fig:FigApp2}d)
whereas the expected integer spike at $U=1$ completely vanishes.

\bibliography{Cuprates}

\end{document}